\documentclass{jfm}

\usepackage{amsmath}
\usepackage{color}
\usepackage{graphicx}
\usepackage{natbib}
\usepackage{overpic}
\usepackage{psfrag}

\ifCUPmtlplainloaded \else
  \checkfont{eurm10}
  \iffontfound
    \IfFileExists{upmath.sty}
      {\typeout{^^JFound AMS Euler Roman fonts on the system,
                   using the 'upmath' package.^^J}%
       \usepackage{upmath}}
      {\typeout{^^JFound AMS Euler Roman fonts on the system, but you
                   dont seem to have the}%
       \typeout{'upmath' package installed. JFM.cls can take advantage
                 of these fonts,^^Jif you use 'upmath' package.^^J}%
      }
  \else
  \fi
\fi

\ifCUPmtlplainloaded \else
  \checkfont{msam10}
  \iffontfound
    \IfFileExists{amssymb.sty}
      {\typeout{^^JFound AMS Symbol fonts on the system, using the
                'amssymb' package.^^J}%
       \usepackage{amssymb}%
         \let\leq=\leqslant
         \let\geq=\geqslant
      }{}
  \fi
\fi

\ifCUPmtlplainloaded \else
  \IfFileExists{amsbsy.sty}
    {\typeout{^^JFound the 'amsbsy' package on the system, using it.^^J}%
     \usepackage{amsbsy}}
    {\providecommand\boldsymbol[1]{\mbox{\boldmath $##1$}}}
\fi

\providecommand\bnabla{\boldsymbol{\nabla}}
\providecommand\bcdot{\boldsymbol{\cdot}}

\newcommand\Rey{\mbox{\textit{Re}}}  

\newsavebox{\astrutbox}
\sbox{\astrutbox}{\rule[-5pt]{0pt}{20pt}}

\newcommand{\ps}   [2]{\ensuremath{\left(\left. {#1} \, \right| \, {#2} \right) }}
\newcommand{\pps}   [2]{\ensuremath{\left(\left(\left. {#1} \, \right| \, {#2} \right)\right) }}

\newcommand\be{\begin{equation}}
\newcommand\ee{\end{equation}}

\def\00{\mathbf{0}}
\def\aa{\mathbf{a}}
\def\AAA{\mathbf{A}}
\def\bb{\mathbf{b}}

\def\bdelta{\boldsymbol{\delta}}
\def\DUs2{\bnabla_{\UU_c} \s2}
\def\ev{\lambda}
\def\evol{\mathbf{P}}
\def\ex{\mathbf{e}_x}

\def\EE{\mathbf{E}}

\def\LL{\mathbf{L}}
\def\MM{\mathbf{M}}
\def\NN{\mathbf{N}}

\def\qq{\mathbf{q}}
\def\qqa{\mathbf{q}^\dag}

\def\s2{G_{opt}^2}
\def\SS{\mathbf{S}}
\def\ua{u^\dag}
 	
\def\uu{\mathbf{u}}
\def\UU{\mathbf{U}}
\def\uua{\mathbf{u}^\dag}		
		
\def\va{v^\dag}
\def\xx{\mathbf{x}}

\definecolor{myorange}{rgb}{0.9,0.4,0}
\definecolor{mygreen}{rgb}{0,0.5,0}
\definecolor{greenstab}{rgb}{.1,.7,.1}
\definecolor{mygrey}{rgb}{.4,.4,.4}

\providecommand\bnabla{\boldsymbol{\nabla}}
\providecommand\bcdot{\boldsymbol{\cdot}}

\title[Second-order sensitivity and optimal spanwise-periodic flow modifications]
{Second-order sensitivity of parallel shear flows and optimal spanwise-periodic flow modifications}

\author[E. Boujo, A. Fani and F. Gallaire]
{
E. Boujo$^{1,2,}$\footnote{Email address for correspondence: edouard.boujo@gmail.com}, 
A. Fani$^1$  and  
F. Gallaire$^1$ 
}

\affiliation{
$^1$ LFMI, \'Ecole Polytechnique F\'ed\'erale de Lausanne, CH-1015 Lausanne, Switzerland\\
$^2$ CAPS, ETH Zurich, CH-8092 Zurich, Switzerland
}

\pubyear{2010}
\volume{650}
\pagerange{119--126}
\date{?; revised ?; accepted ?. - To be entered by editorial office}
\begin{document}

\maketitle

\begin{abstract}
The question of optimal spanwise-periodic  modification for the stabilisation of  spanwise-invariant flows is addressed. A second-order sensitivity analysis is conducted for the linear temporal stability of parallel flows $U_0$ subject to small-amplitude spanwise-periodic modification $\epsilon U_1$, $\epsilon \ll 1$. It is known that spanwise-periodic flow modifications have a quadratic effect on stability properties, i.e. the first-order eigenvalue variation is zero, hence the need for a second-order analysis. A second-order sensitivity operator is computed from a one-dimensional calculation, which allows one to predict how eigenvalues are affected by any flow modification $U_1$, without actually solving for modified eigenvalues and eigenmodes. Comparisons with full two-dimensional stability calculations in a plane channel flow and in a mixing layer show excellent agreement. 

Next, optimisation is performed on the second-order sensitivity operator: for each eigenmode streamwise wavenumber $\alpha_0$ and  base flow modification spanwise wavenumber $\beta$, the most stabilising/destabilising profiles $U_1$ are computed, together with lower/upper bounds for the variation in leading eigenvalue. These bounds increase like $\beta^{-2}$ as $\beta$ goes to zero, thus yielding a large stabilising potential. However, three-dimensional modes with wavenumbers $\beta_0=\pm\beta$, $\pm\beta/2$ are destabilised, and therefore larger control wavenumbers should be preferred. The most stabilising $U_1$ optimised for the most unstable streamwise wavenumber $\alpha_{0,max}$ has a stabilising effect on modes with other $\alpha_0$ values too. 

Finally, the potential of transient growth to amplify perturbations and stabilise the flow is assessed with a combined optimisation. Assuming a separation of time-scales between the fast unstable mode and the slow transient evolution of the optimal perturbations, combined optimal perturbations that achieve the best balance between transient linear amplification and stabilisation of the nominal shear flow are determined. In the mixing layer with $\beta\leq1.5$, these combined optimal perturbations appear similar to transient growth-only optimal perturbations, and achieve a more efficient overall stabilisation than optimal spanwise-periodic and spanwise-invariant modifications computed for stabilisation only. These results are consistent with the efficiency of streak-based control strategies. 
\end{abstract}


\section{Introduction}

It is well known that the addition of three-dimensional (3D) spanwise-periodic perturbations to a two-dimensional (2D) cylinder wake can stabilise vortex shedding \citep{Zdravkovich81, Kim05, Choi08}. 
Recently, \cite{Hwang13}  proposed  an explanation of this effect based on the linear stability  of the spanwise-modulated parallel wake flow. 
They demonstrated a substantial attenuation of the absolute instability growth rate in a range of wavelengths  corresponding to results from experiment and direct numerical simulation.

In a recent paper, \cite{DelGuercio2014}  showed that suitable spanwise-periodic perturbations added to nominally parallel wakes significantly reduce the temporal stability growth rate of the inflectional instability and completely quench the absolute instability as well. 
Perturbations to the nominally parallel base flow were chosen as  non-linear streaks resulting from the  optimal lift-up mechanism, i.e. the transient  
growth of optimal streamwise-uniform spanwise-periodic vortices. 
The  absolute and maximum temporal growth rates  were found to depend quadratically on the streak amplitudes, as suggested by \cite{Hwang13}  who demonstrated that the  linear sensitivity was zero for spanwise-periodic  disturbances of the base flow.

\cite{Cossu14secondorder} outlined a rigorous mathematical procedure to compute beforehand the quadratic (second-order) sensitivity of an eigenmode when its linear (first-order) sensitivity vanishes. He used this method to explain the stabilisation of global modes of the one-dimensional (1D) Ginzburg-Landau equation, serving as a model equation for spatially developing shear flows submitted to spanwise-periodic modulations.
\cite{Tammi14} applied a similar technique to investigate the stabilisation of the 2D wake behind a flat plate using spanwise-periodic wall actuation.
In these two studies, \textit{particular} base flow modifications were prescribed  and the resulting first-order eigenmode variation had to be computed explicitly to obtain the eigenvalue variation.

In this paper, we address the question of \textit{optimal} spanwise-periodic flow modification.
In a first step, we use an asymptotic expansion  to express the second-order eigenvalue variation, and we determine a second-order sensitivity operator which allows us to predict the eigenvalue variation resulting from \textit{any} base flow modification without ever computing the first-order eigenmode correction and without solving the modified eigenvalue problem. 
Then, we optimise the eigenvalue variation and obtain \textit{optimal} flow modifications, i.e. the optimally destabilising/stabilising  spanwise-periodic base flow modulations.
We illustrate this optimisation technique with the classical hydrodynamic stability of nominally parallel flows governed by the linearised Navier--Stokes equations at finite  Reynolds number $\Rey$ (the so-called Orr--Sommerfeld equations) in the unstable regime: the plane channel flow and a prototypical  mixing layer.
A normal mode expansion yields a set of eigenvalues and eigenmodes for any given streamwise and spanwise wavenumbers $(\alpha_0,\beta_0)$, and in both flows the most unstable mode is two-dimensional ($\beta_0=0$). 
We consider  base flow modifications  that are unidirectional and parallel to the base flow direction, and spanwise-periodic of wavenumber $\beta$. 
For each  wavenumber pair $(\alpha_0,\beta)$,
we optimise the variation of the leading 2D eigenvalue,
thus obtaining bounds for this variation together with the associated optimally destabilising/stabilising base flow modifications. 
The effect on other modes $(\alpha_0,\beta_0)$ is also discussed.

In a second step, we incorporate transient growth in the optimisation procedure and find \textit{combined optimal} perturbations
that achieve the best trade-off between transient amplification and stabilisation. 
Indeed, flow modifications optimised for pure  stabilisation do not take advantage of non-normal amplification mechanisms and may thus require a substantial amplitude; on the other hand, perturbations optimised for pure transient growth may not be well suited to stabilisation.
\textcolor{black}{
This combined optimisation is conducted under the strong hypothesis of a separation of time scales between the fast instability  of the nominal base flow and the slow evolution of purely transverse steady initial perturbations into streaks \citep{Reddy98, Cossu04}.
}

Section \ref{sec:problem}  details  the derivation of the second-order sensitivity operator and the optimisation method.
Sections \ref{sec:planechannel}-\ref{sec:mixinglayer} are devoted to optimal flow modifications in the plane channel flow and in the mixing layer. 
Section \ref{sec:combined} presents the combined optimisation of transient growth and flow modification, and characterises the  performance of optimal streaks in different $\beta$ ranges. Comparison of spanwise-uniform and spanwise-periodic optimals is also provided.

%
\section{Problem formulation}
\label{sec:problem}

Given a steady base flow $(\UU,P)^T$ solution of the Navier--Stokes (NS) equations, the dynamics of small-amplitude perturbations $\hat\qq=(\hat\uu,\hat p)^T=(\hat u,\hat v,\hat w,\hat p)^T$ superimposed onto this base flow are governed by the linearised NS equations
\be 
\partial_t \hat\uu + \UU\bcdot\bnabla \hat\uu  
+ \hat\uu\bcdot\bnabla \UU 
+ \bnabla  \hat p -\Rey^{-1}\bnabla^2 \hat\uu = \00,
\quad
\bnabla \bcdot \hat\uu = 0.
 \label{eq:LNS}
\ee

\subsection{Second-order eigenvalue variation}

Consider a  1D parallel flow $U_0(y){\bf e}_x$ perturbed with a 2D spanwise-periodic modification of small amplitude $\epsilon$:
\begin{equation}
{\bf U}(y,z)= \left( U_0 (y)+\epsilon U_1(y)\cos(\beta z) \right) {\bf e}_x.
\label{eq:U0U1}
\end{equation}
Assuming a normal mode expansion
$\hat \qq(x,y,z,t)=\qq(y,z)\exp{(i\alpha_0x+\ev t)}$
in spanwise coordinate $x$ 
and time $t$ for small-amplitude perturbations 
and linearising the NS equations results in the eigenvalue problem  
\be 
\lambda \uu + \UU\bcdot\bnabla \uu  
+ \uu\bcdot\bnabla \UU 
+ \bnabla  p -\Rey^{-1}\bnabla^2 \uu = \00,
\quad
\bnabla \bcdot \uu = 0
 \label{eq:LNS_evp}
\ee
for the eigenvalue $\ev=\ev_r +i \ev_i$.
The set of growth rates $\ev_r$ and frequencies $\ev_i$
determines the linear temporal stability properties of the flow.

\begin{figure}
  \centerline{
  	\begin{overpic}[height=5cm,tics=10]{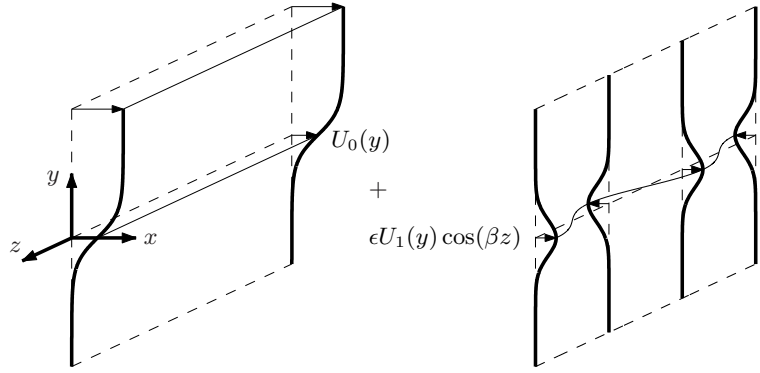}
  	\put(17,17.) {$x$}
  	\put( 4,  25.5) {$y$}
  	\put(-1,  16.) {$z$}
  	\put(42,30) {\textcolor{black}{$U_0(y)$}}
  	\put(47,23.5) {+}
  	\put(47,17) {\textcolor{black}{$\epsilon U_1(y) \cos(\beta z)$}}
   	\end{overpic}
  }    
   \caption{Sketch of the base flow configuration: a 1D parallel flow $U_0(y)$  is modified with a small-amplitude 2D spanwise-periodic flow  $\epsilon U_1(y) \cos(\beta z)$.}
\label{fig:sketch}
\end{figure}

While it is common at this stage, for spanwise-invariant parallel base flows,  to transform the primitive variables $\qq=(\uu,p)^T$ into the Orr-Sommerfeld variables 
(cross-stream velocity and vorticity), we keep here the primitive variables  and write  (\ref{eq:LNS_evp}) formally as 
$\ev \bf E \qq+{\bf A}\qq=\00$.
Since the linearised NS
operator ${\bf A}$ 
is linear in the base flow ${\bf U}$, it can be expanded exactly to arbitrary order (here to second order) as ${\bf A}={\bf A}_0+\epsilon{\bf A}_1$. 
The singular matrix ${\bf E}$ need not be expanded since it does not depend on the base flow. Detailed expressions for all  operators are given in
Appendix~\ref{sec:app_stab}. 
It should be noted in particular that $\AAA_0$ depends on the unperturbed base flow $U_0$, and $\AAA_1$ depends on the  base flow modification $U_1$.

We look for perturbed eigenvalues and eigenmodes using the following 
expansion in the amplitude $\epsilon$ of the base flow modification:
\begin{equation}
\lambda = \lambda_0+\epsilon \lambda_1+\epsilon^2 \lambda_2+\ldots, 
\quad
{\bf q} = {\bf q}_0+\epsilon {\bf q}_1 +\epsilon^2 {\bf q}_2 + \ldots
\label{eq:exp}
\end{equation}
By substituting (\ref{eq:exp}) into (\ref{eq:LNS_evp}),  we recover at leading order ($\epsilon^0$) the linearised  NS  equations
\be
\lambda_0 \uu_0 + \UU_0\bcdot\bnabla \uu_0  + \uu_0\bcdot\bnabla \UU_0 + \bnabla  p_0 -\Rey^{-1}\bnabla^2 \uu_0  = \00,
\quad \bnabla \bcdot \uu_0 = 0,
\label{eq:EVP0}
\ee
which we write as an eigenvalue problem for the eigenvalue $\lambda_0$ and eigenmode ${\bf q}_0$:
\begin{equation}
(\lambda_0 {\bf E}+{\bf A}_0) {\bf q}_0=\bf 0.
\label{eq:evp1111}
\end{equation}
In the following we focus on the leading (most unstable) eigenvalue and its associated eigenmode, which is 2D for inflectional velocity profiles: $\partial_z \qq_0=\00, w_0=0$.

At first order ($\epsilon^1$) we obtain
\begin{equation}
  (\lambda_0 {\bf E} +{\bf A}_0){\bf q}_1
+ (\lambda_1 {\bf E} +{\bf A}_1){\bf q}_0 = \bf 0.
\label{eq:evp1}
\end{equation}
We introduce the following 1D and 2D Hermitian inner products
\be 
\displaystyle \ps{\aa}{\bb} =  \lim_{L_y \rightarrow \infty} 
\int_{-L_y/2}^{L_y/2}\overline\aa\bcdot\bb \,\mathrm{d}y,
\quad
\displaystyle \pps{\aa}{\bb} = \lim_{L_z \rightarrow \infty} 
\frac{1}{L_z}
\int_{-L_z/2}^{L_z/2} \ps{\aa}{\bb} \,\mathrm{d}z,
\ee
where the overbar stands for complex conjugate.
For any operator $\NN$ we denote $\NN^\dag$ the adjoint operator such that
$\pps{\aa}{\NN\bb}=\pps{\NN^\dag\aa}{\bb}$ $\forall \, \aa,\bb$.
We  obtain the first-order eigenvalue variation $\ev_1$ by projecting (\ref{eq:evp1}) on the leading adjoint eigenmode $\qqa_0$ defined by
$(\overline\ev_0 \EE + \AAA_0^\dag) \qq_0^\dag = \00$
and  normalised with $\ps{ {\bf q}_0^\dag } {\EE {\bf q}_0}=1$ \citep{Hinch91, Trefethen93, Chomaz05, Giannetti07}:
\be 
\lambda_1  = - \pps{ {\bf q}_0^\dag } {{\bf A}_1{\bf q}_0}.
\label{eq:project_adjoint}
\ee
Since  ${\bf A}_1$ is periodic in $z$, the inner product in (\ref{eq:project_adjoint}) vanishes and
the first-order eigenvalue variation is zero:
 $\lambda_1=0$. 
In other words, spanwise-periodic flow modifications have no first-order effect on stability properties \citep{Hwang13, DelGuercio2014,Cossu14secondorder}.
From (\ref{eq:evp1}), the leading eigenmode variation is 
$\qq_1 = -(\ev_0\EE+\AAA_0)^{-1}\AAA_1\qq_0$. 
In this expression the operator $(\ev_0\EE+\AAA_0)$  is not invertible in general since (\ref{eq:evp1111}) has a non-trivial solution, but the inverse is taken in the subspace orthogonal to $\qq_0$, and $\qq_1$ is defined up to any constant component in the direction of $\qq_0$ \citep{Hinch91}.
This is made possible by the solvability condition (Fredholm theorem) for (\ref{eq:evp1}) to be satisfied:
the forcing term $(\lambda_0 {\bf E} +{\bf A}_1){\bf q}_0={\bf A}_1{\bf q}_0$ is orthogonal to the solution $\qqa_0$ of  the  adjoint equation 
associated with (\ref{eq:evp1111}), as expressed precisely by (\ref{eq:project_adjoint}).

At second order ($\epsilon^2$)   we obtain
$ (\lambda_0 {\bf E}+{\bf A}_0){\bf q}_2
+{\bf A}_1{\bf q}_1
+ \lambda_2 {\bf E}{\bf q}_0 = \bf 0,$
which yields after projection on the leading adjoint eigenmode
\begin{equation}
 \ev_2 = -\pps{\qq_0^\dag}{\AAA_1\qq_1}
 = \pps{\qq_0^\dag}{\AAA_1 (\ev_0\EE+\AAA_0)^{-1}\AAA_1\qq_0}.
\label{eq:ev2}
\end{equation}

For a \textit{given} flow modification $U_1$, one can explicitely compute the modification $\AAA_1$ of the linearised NS operator and use  (\ref{eq:ev2}) to obtain the second-order eigenvalue variation.
However, in order to investigate the effect of \textit{any} flow modification, it is desirable to manipulate this expression to isolate a sensitivity operator independent of $U_1$, similar to classical first-order sensitivity analyses \citep{Hill92AIAA, Marquet08cyl, Meliga10}.
We use adjoint operators to perform this manipulation in section~\ref{sec:sensit}.

%
\subsection{Second-order sensitivity}
\label{sec:sensit}

We look for a sensitivity operator $\SS_2$ such that the second-order eigenvalue variation induced by any small spanwise-periodic flow modification $(U_1(y)\cos(\beta z),0,0)^T$ as given in (\ref{eq:U0U1}) is easily predicted by the inner product 
\begin{equation}
 \ev_2 = \pps{ U_1}{\SS_2 U_1}.
\end{equation}
We rewrite (\ref{eq:ev2}) as 
$\ev_2 = \pps{\AAA_1^\dag \qq_0^\dag}{ (\ev_0\EE+\AAA_0)^{-1}\AAA_1\qq_0}$
and, given that $\AAA_1$ is linear in $U_1$, we introduce operators $\MM$ and $\LL$ such that 
$\AAA_1^\dag\qqa_0=\MM U_1$
and 
$\AAA_1\qq_0=\LL U_1$ (detailed expressions are given in 
Appendix~\ref{sec:app_sensit}).
We finally obtain
\be 
 \ev_2 
= \pps{ U_1}{ \MM^\dag(\ev_0\EE+\AAA_0)^{-1} \LL U_1}
\label{eq:ev21}
\ee

At this point $\SS_2=\MM^\dag(\ev_0\EE+\AAA_0)^{-1} \LL$ depends on $z$, but 
(\ref{eq:ev21}) only contains terms  proportional to
 $\cos^2(\beta z)$ and $\sin^2(\beta z)$ and can therefore be replaced with
\begin{equation}
 \ev_2
 =  \ps{ U_1}{ \widetilde\SS_2 U_1}
 =  \ps{ U_1}{\frac{1}{2} \widetilde\MM^\dag(\ev_0\EE + \widetilde\AAA_0)^{-1} \widetilde \LL U_1}
\label{eq:ev23}
\end{equation}
where $\widetilde\MM^\dag$ and $\widetilde \LL$  are $z$-independent 
 versions of $\MM^\dag$ and $\LL$, 
and $z$-derivatives in $\AAA_0$ are appropriately replaced with $\beta$ terms in $\widetilde\AAA_0$ 
(see also Appendix~\ref{sec:app_z}):
\be
\widetilde
\MM^\dag
 = \left[
 i \alpha_0 \overline u_0^\dag,\,
   i \alpha_0 \overline v_0^\dag 
  -\partial_y\overline u_0^\dag
  -\overline u_0^\dag\partial_y,\,
-\beta \overline u_0^\dag,\, 
0 
\right],
\quad
\widetilde \LL
= \left[
 i \alpha_0 u_0 +  v_0 \partial_y,\, 
  i \alpha_0 v_0,\, 
0,\, 
0
\right]^T,
\ee
\begin{equation}
\widetilde{\bf A}_0
=
\left[\begin{array}{cccc}
i\alpha_0U_0-\Rey^{-1} \nabla_{\alpha_0\beta} & \partial_y U_0 & 0 & i\alpha_0\\
0 & i\alpha_0U_0-\Rey^{-1}  \nabla_{\alpha_0\beta} & 0 & \partial_y \\
0 & 0 & i\alpha_0U_0-\Rey^{-1}  \nabla_{\alpha_0\beta} &  -\beta\\
i \alpha_0 & \partial_y & \beta & 0 
\end{array}\right],
\end{equation}
$\nabla_{\alpha_0\beta} = -\alpha_0^2+\partial_{yy} -\beta^2$,
yielding the $z$-independent  sensitivity 
 $\widetilde\SS_2 = \frac{1}{2} \widetilde\MM^\dag(\ev_0\EE + \widetilde\AAA_0)^{-1} \widetilde \LL $.

An important difference from first-order sensitivity analysis is that $\ev_1$ ($\neq 0$ in  general) depends linearly on $U_1$,  while here $\ev_2$ depends quadratically on $U_1$.
As a consequence, in the case of general first-order sensitivity, one can investigate the effect of any base flow modification $\UU_1$ in a very convenient way: since $\ev_1$ can be written $\pps{\SS_1}{\UU_1}$ and is linear in $\UU_1$, a linear combination of  flow modifications results in the same linear combination of eigenvalue variations. In particular, any $\UU_1$ can be decomposed into a combination of flow modifications  localised in $\xx_c$ (e.g. Gaussian functions approximating pointwise Dirac delta functions), each of them resulting in an individual eigenvalue variation obtained from the sensitivity at $\xx_c$ only;
since $\SS_1$ is a vector field, it is easily visualised with a map, and one can  identify the most sensitive regions at a glance \citep{Marquet08cyl, Meliga10}.

In contrast, in the general case of second-order sensitivity, $\ev_2=\pps{\UU_1}{\SS_2 \UU_1}$ depends quadratically on $\UU_1$, and a linear combination of base flow modifications does \textit{not} result in the same combination of eigenvalue variations, not to mention quadratic coupling effects between different components $U_1$, $ V_1$, $W_1$ (as appears clearly in the alternative expression $\ev_2=(\UU_1\UU_1^T ): \SS_2$
 denoting the inner product of $\SS_2$ with the tensor $\UU_1 \UU_1^T$).
Furthermore, the sensitivity operator $\SS_2$ is a tensor field, whose visualisation would require an impractically large number of maps.
\cite{Tammi14} proposed to  identify sensitive regions  with maps showing the effect of a specific base flow modification (localised Gaussian $U_1$, and $V_1=W_1=0$) applied successively in all locations of the domain, essentially reproducing systematic traversing experimental measurements.
We prefer to take advantage of  knowing the sensitivity operator $\SS_2$,
and we show in section~\ref{sec:optimal} how to extract \textit{optimal} flow modifications resulting in maximal eigenvalue variation.

\subsection{Optimal flow modification}
\label{sec:optimal}

The second-order sensitivity operator is useful in that it predicts the leading eigenvalue variation resulting from any $U_1$ without the need to solve the eigenvalue problem for the modified flow.
In addition, it allows one to determine the largest possible eigenvalue variation for all modifications of given amplitude,
i.e. maximise/minimise $\ev_2$.
With the eigenvalue variation recast as (\ref{eq:ev23}),
this optimisation problem is equivalent to an eigenvalue problem, formally similar to optimal transient growth and optimal harmonic response.
Specifically, the largest increase (decrease) in growth rate to be expected for a modification of unit norm 
$||U_1||=\ps{U_1}{U_1}^{1/2}=1$
is the largest positive (largest negative) eigenvalue of the symmetric real part of $\widetilde \SS_2$:
\begin{subeqnarray}
& \displaystyle \max_{||U_1||=1} \ev_{2r} 
=   \max_{||U_1||=1} 
\ps{U_1} { \frac{1}{2} \left( \widetilde\SS_{2r}+\widetilde\SS_{2r}^T \right) U_1}
= \ev_{max}
\left\{ 
\frac{1}{2}
\left( \widetilde\SS_{2r}+\widetilde\SS_{2r}^T \right)
\right\},
\label{eq:mostdestab}
\\
& \displaystyle \min_{||U_1||=1} \ev_{2r} 
=   \min_{||U_1||=1} 
\ps{U_1} { \frac{1}{2} \left( \widetilde\SS_{2r}+\widetilde\SS_{2r}^T \right) U_1}
=  \ev_{min}
\left\{ 
 \frac{1}{2}
 \left( \widetilde\SS_{2r}+\widetilde\SS_{2r}^T \right)
\right\},
\label{eq:moststab}
\end{subeqnarray}
where $\widetilde \SS_2 = \widetilde\SS_{2r}+i \widetilde\SS_{2i}$,
and the right-hand sides come from
$\widetilde\SS_{2r}+\widetilde\SS_{2r}^T$ 
being real symmetric, so that the Rayleigh quotient is maximal (minimal) for the largest positive (largest negative) eigenvalue.
The optimal flow modification, i.e. the most stabilising (destabilising) $U_1$ is the eigenmode of unit norm associated with $\ev_{max}$ ($\ev_{min}$).
Similar relations hold for the  maximal shift in frequency, real parts  being replaced with imaginary parts.
If needed, one can also solve for other eigenvalues and obtain a sequence of orthogonal suboptimal flow modifications associated with smaller values of $|\ev_{2r}|$.

\section{Results: the plane channel}
\label{sec:planechannel}

We investigate the  parallel flow in a plane channel between solid walls located at $y=\pm 1$.
The base flow has a Poiseuille parabolic profile $U_0(y)=1-y^2$ and first becomes linearly unstable at $\Rey=5772$.
We solve the eigenvalue problem (\ref{eq:LNS_evp}) in primitive variables $(\uu,p)^T$ with a 1D spectral method using $N=100$ Chebyshev polynomials \citep{Trefethen00Matlab} on $y\in[-1;1]$ and  homogeneous Dirichlet boundary conditions on velocity components.
The method was validated against spectral results in Orr-Sommerfeld variables \citep{Schmid01},  yielding $10^{-8}$ accuracy with $N=35$ polynomials.
We focus on the Reynolds number $\Rey=6000$, where the largest growth rate is obtained for wavenumbers $(\alpha_0,\beta_0)=(\alpha_{0,max},\beta_{0,max})=(1.016,0)$, and the leading eigenvalue  
$\ev_0 = 3.694\times 10^{-4} -0.2659 i$  corresponds to a 2D Tollmien–-Schlichting wave.

\subsection{Optimal flow modifications}
\label{sec:POIS_optimflowmodif}

\begin{figure}
  \psfrag{beta}[t][]{$\beta$}
  \psfrag{alpha}[t][]{$\alpha_0$}
  \psfrag{lam2}   [][]{} 	
  \psfrag{lam2max}[b][]{}
  \psfrag{lam2ds}[b][]{} 	
  \psfrag{betamax}[b][]{} 	
  \vspace{0.8cm}
  \centerline{
  	\begin{overpic}[height=5.5cm,tics=10]{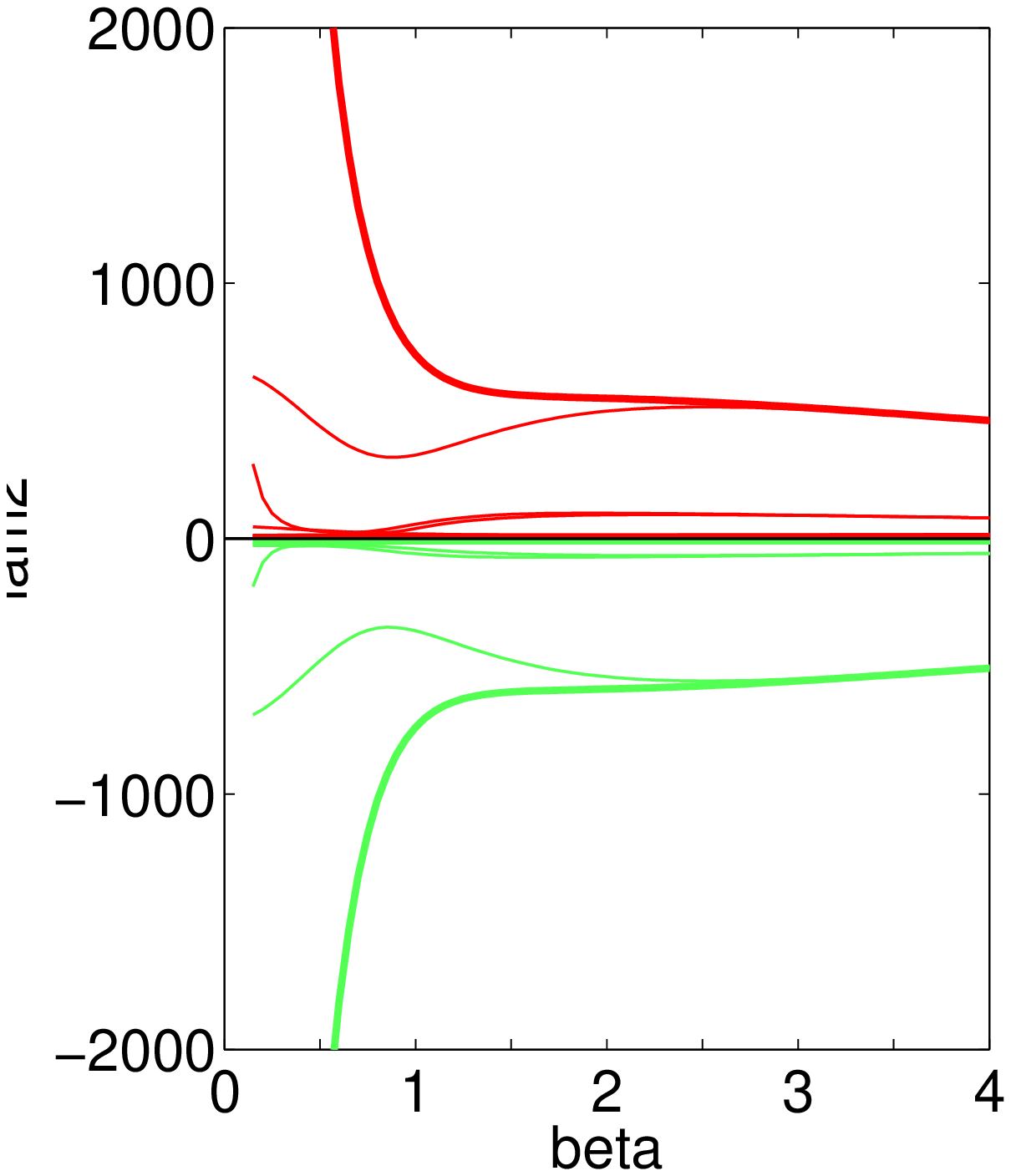}
	  	\put(18,103){$(a)$ $\min \ev_{2r}$, $\max \ev_{2r}$}		
  		\put(32,77){\textcolor{red}{I$^d$}}
  		\put(20,70){\textcolor{red}{II$^d$}} 
  		\put(20,34){\textcolor{greenstab}{II$^s$}}
  		\put(32,25){\textcolor{greenstab}{I$^s$}}
   	\end{overpic}
   	\hspace{0.2cm}
  	\begin{overpic}[height=5.53cm,tics=10]{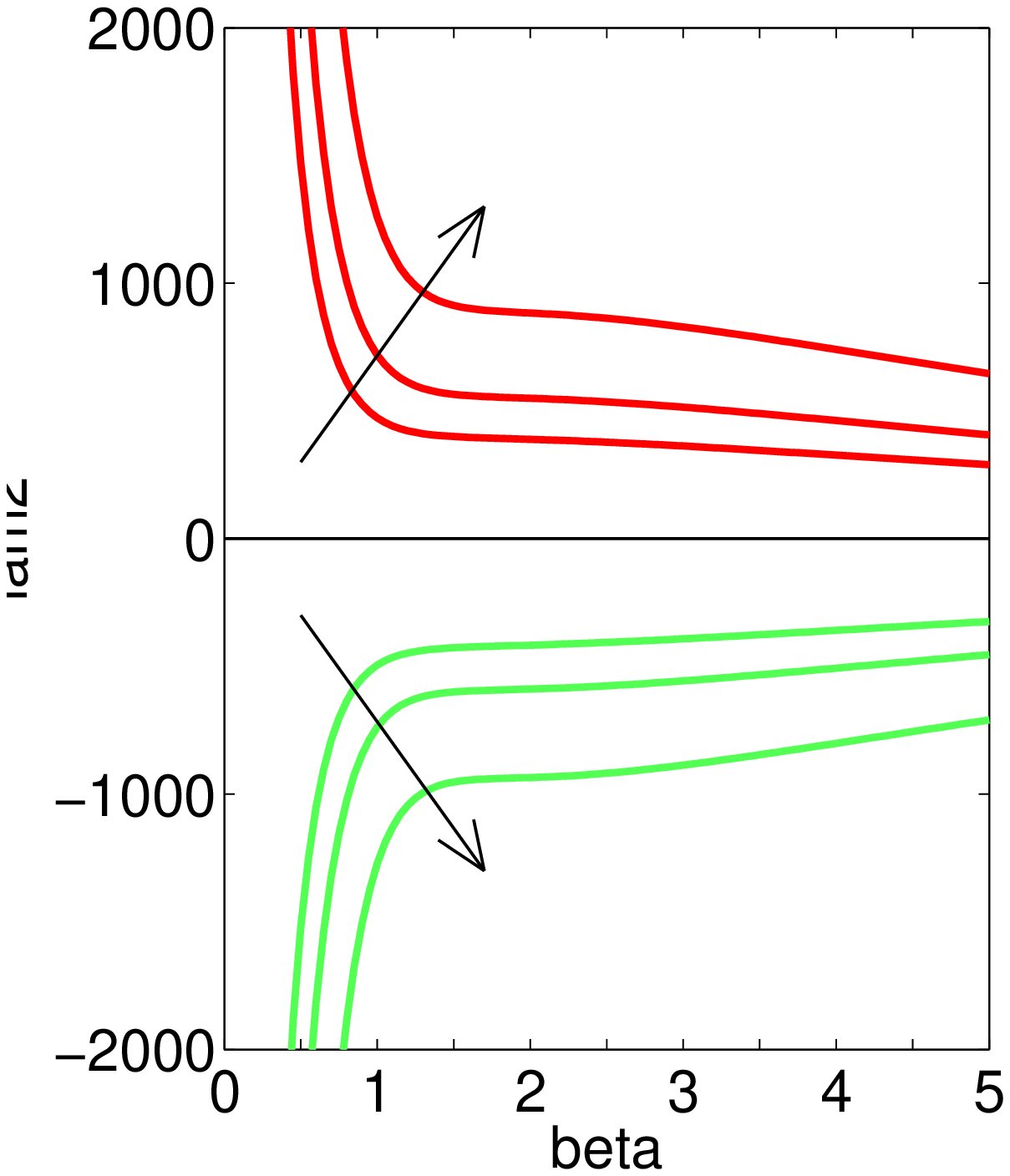}
	  	\put(18,102.5){$(b)$ $\min \ev_{2r}$, $\max \ev_{2r}$} 
	  	\put(37,85){\footnotesize $\alpha_0=0.9, 1.0, 1.1$}  
   	\end{overpic}
  }   
  \vspace{-0.01cm} 
  \caption{Variation of the leading growth rate for optimally destabilising flow modification (superscript $d$, dark lines, red online) and optimally stabilising flow modification (superscript $s$, light lines, green online), as predicted by sensitivity analysis. 
   $(a)$~Optimisation for the leading eigenmode of wavenumbers $(\alpha_0,\beta_0)=(1,0)$. suboptimals are also shown as thin lines.
   $(b)$~Optimal growth rate variation for $\alpha_0=0.9, 1.0, 1.1$ and $\beta_0=0$.
 }
\label{fig:POIS_lam_vs_beta}
\end{figure}

We compute the most destabilising and most stabilising spanwise-periodic flow modifications according to (\ref{eq:mostdestab}).
Figure~\ref{fig:POIS_lam_vs_beta}$(a)$ shows the largest positive and negative second-order eigenvalue variations at
$(\alpha_0,\beta_0)=(1,0)$ as a function of the control spanwise wavenumber $\beta$.
The curves labelled I$^d$ and I$^s$ provide bounds for the largest possible destabilisation and stabilisation. 
As shown in figure~\ref{fig:POIS_lam_vs_beta}$(b)$,
the upper and lower bounds increase with $\alpha_0$ 
indicating that optimal flow modifications have a stronger effect on 
leading eigenmodes of smaller streamwise wavelengths.
It should be noted that  $\min \ev_{2r}$  and $\max \ev_{2r}$ are of the same order of magnitude in absolute value.

Figure~\ref{fig:POIS_lam_vs_beta}$(a)$ also shows growth rate variations for suboptimal flow modifications (thin lines). 
The first suboptimal modifications (branches II$^d$, II$^s$) become as effective as the optimal ones when $\beta \geq 2.5$.

\begin{figure}
  \def\thisfigy{80} 
  \psfrag{yy} [r][][1][-90]{$y$}	
  \psfrag{y}  [r][][1][-90]{$y$}	
  \psfrag{UU1}[t][]{$U_1$}	
  \psfrag{U1} [t][]{$U_1$}
  \vspace{0.5cm}
  \centerline{
    \hspace{0.5cm}
  	\begin{overpic}[height=4.9cm,tics=10]{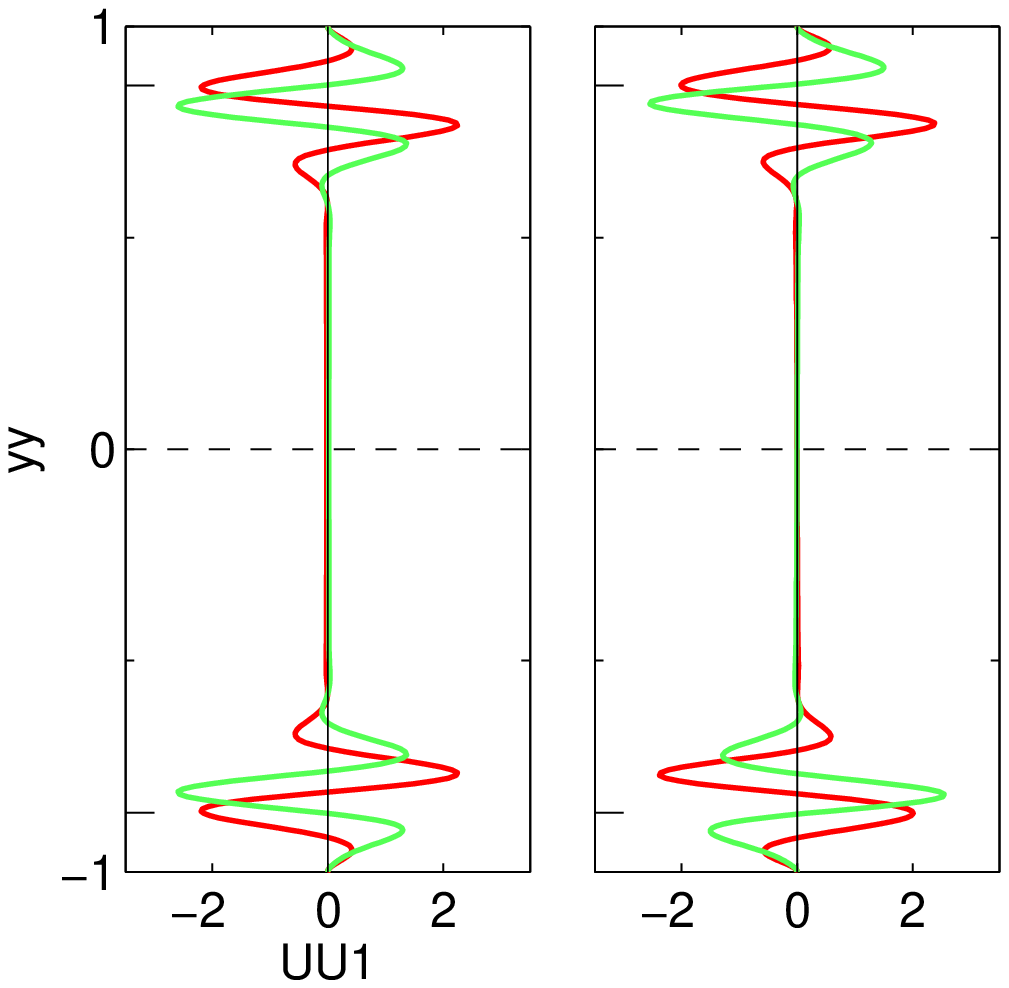}
   		\put(-15,90){$\beta=0.4$}
  		\put(25,96){\textcolor{red}{I$^d$}  $\quad$ \textcolor{greenstab}{I$^s$}}
  		\put(67,96){\textcolor{red}{II$^d$} $\quad$ \textcolor{greenstab}{II$^s$}}
   	\end{overpic}
    \hspace{1.2cm}
   	\begin{overpic}[height=4.9cm,tics=10]{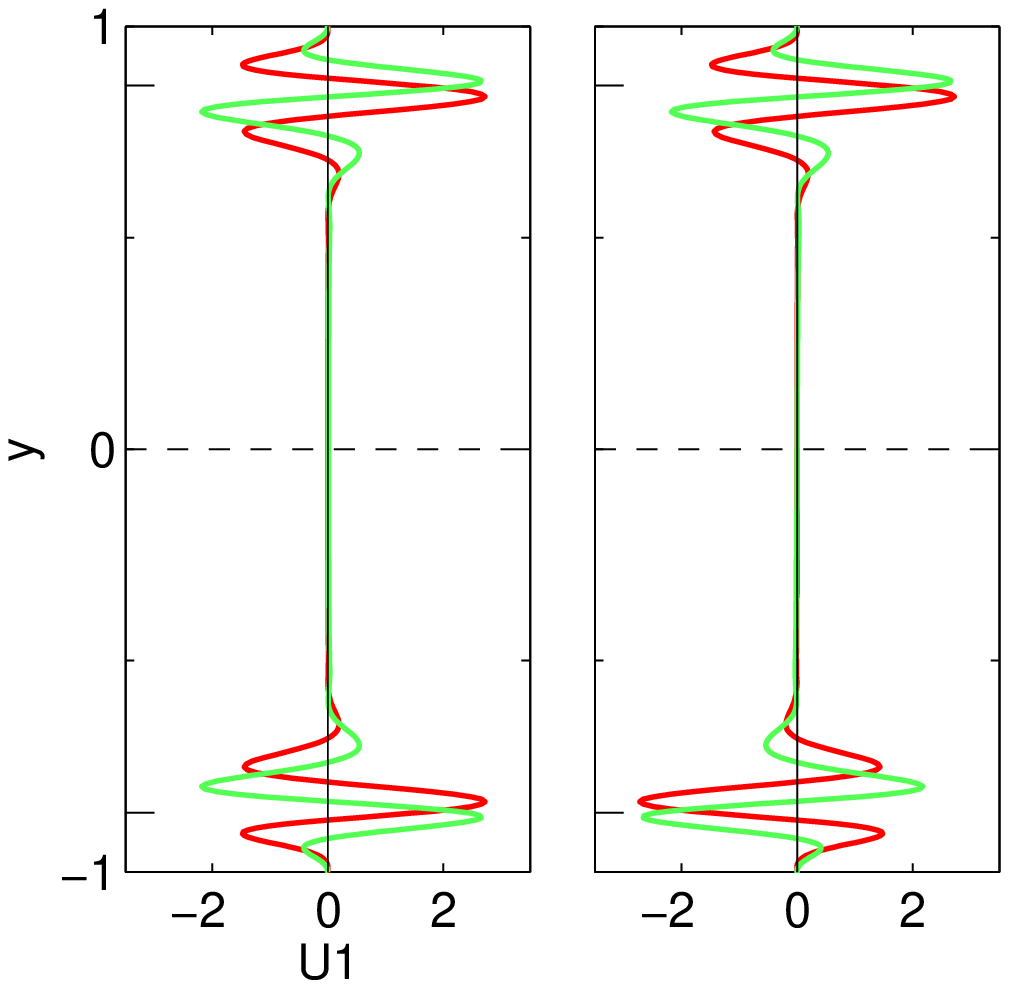}
  		\put(-15,90){ $\beta=3$}
  		\put(25,96){\textcolor{red}{I$^d$}  $\quad$ \textcolor{greenstab}{I$^s$}}
  		\put(67,96){\textcolor{red}{II$^d$} $\quad$ \textcolor{greenstab}{II$^s$}}
   	\end{overpic}  
  }   
   \caption{Optimal (branch I) and first suboptimal (branch II)
 destabilising (dark, red online) and stabilising (light, green online) base flow modification for the most unstable mode $(\alpha_0,\beta_0)=(\alpha_{0,max},\beta_{0,max})=(1,0)$, at control wavenumbers $\beta=0.4$ and $\beta=3$.
Long ticks on the left show the position of the critical layers at $y=\pm0.86$.}
\label{fig:POIS_U0}
\end{figure}

For all  wavenumbers $\beta$, the optimal stabilising and destabilising  modifications  for $(\alpha_0,\beta_0)=(1,0)$, shown in figure~\ref{fig:POIS_U0}, are symmetric in $y$,
and are localised close to the critical layer,
where the base flow velocity is equal to the phase speed of the Tollmien--Schlichting wave ($y=\pm0.86$).
The first suboptimals have similar structures but are antisymmetric.

\begin{figure}
  \psfrag{beta}[t][]{$\beta$}
  \psfrag{alpha}[t][]{$\alpha_0$}
  \psfrag{abs(lam2)}[r][][1][-90]{$|\min \ev_{2r}|$, $|\max \ev_{2r}|$}	
  \psfrag{lam2max}[b][]{}
  \psfrag{lam2ds}[b][]{} 
  \psfrag{betamax}[b][]{} 	
  \centerline{
  	\begin{overpic}[height=5.46cm,tics=10]{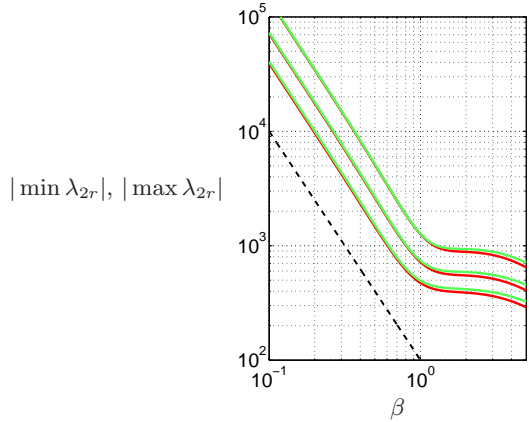}
   	\end{overpic}
  }   
  \caption{Optimal eigenvalue variations  diverge like $\beta^{-2}$ as the control  wavenumber decreases 
(same data as figure~\ref{fig:POIS_lam_vs_beta} in logarithmic scale, $\alpha_0=0.9, 1.0, 1.1, \beta_0=0$). }
\label{fig:POIS_beta2_slope}
\end{figure}

For small values of $\beta$, the upper and lower bounds diverge like $\beta^{-2}$,
as shown in figure~\ref{fig:POIS_beta2_slope}.
This points to a strong authority of the flow modification as its spanwise wavelength tends to infinity.
This diverging effect at small $\beta$ was also observed by \cite{DelGuercio2014} and \cite{Cossu14secondorder} 
 for a specific choice of base flow modification (streaks created by optimal streamwise vortices in a 2D wake, and uniform modification in a 1D Ginzburg-Landau equation respectively). 
Based on an expansion on the eigenmodes of the unperturbed problem \citep{Hinch91}, this phenomenon was explained by \cite{Tammi14} as a modal resonance between unperturbed eigenmodes having close eigenvalues, namely  the eigenmode of interest (2D mode $\beta_0=0$ in our case) and eigenmodes whose spanwise wavenumbers differ by $\pm \beta$ from that of the eigenmode of interest (i.e. $\beta_0=\pm \beta$). 
We show in Appendix~\ref{sec:app_scaling} that the eigenvalue difference between the 2D eigenmode of interest and  eigenmodes of small spanwise wavenumber $\beta_0=\pm \beta$ scales precisely like $\beta^{2}$, consistent with the divergence behaviour $\ev_2  \sim \beta^{-2}$ we observe.
It should be mentioned in this context \citep{Tammi14} that if $\beta$ is small enough  that the eigenvalue difference  between these modes is small and of order $\Delta \ev \sim \epsilon$, then  the modal resonance results in a non-small second-order eigenvalue variation $\ev_2=\mathcal{O}(\epsilon/\Delta \ev)=\mathcal{O}(1)$ and the perturbation approach (\ref{eq:exp}) is not valid any longer. 
Experimentally, the spanwise extension of the system sets a minimal value for $\beta$.

\subsection{Validation}
\label{sec:POIS_valid}

In order to validate the sensitivity analysis, full 2D calculations were performed on base flows with optimally stabilising spanwise-periodic modifications.
The numerical method is the following.
Equations (\ref{eq:LNS_evp}) are discretised on the computational domain   $y\in[-1;1]$, $z\in[0; 5\cdot 2\pi/\beta]$ with the finite element solver FreeFem++ \citep{freefem} using P2 elements for  velocity components and P1 elements for pressure; 
homogeneous Dirichlet boundary conditions $\mathbf{u}=\mathbf{0}$ are imposed at $y =\pm 1$, while  periodic boundary conditions are imposed on  lateral boundaries ($z=0$, $z=5\cdot 2\pi/\beta$); finally the library SLEPc \citep{slepc} is used to compute generalised eigenpairs of the eigenvalue problem with base flow $U_0(y)+\epsilon U_1(y) \cos(\beta z)$, and $U_1$ of unit 1D norm.

\begin{figure}
  \psfrag{k=1}[t][]{}
  \psfrag{U}[t][]{$U_1$}
  \psfrag{eps}[t][]{$\epsilon$}
  \psfrag{y}[][][1][-90]{$y$}
  \psfrag{real(lambda0)}[][][1][-90]{$\ev_{r}\quad$}
  \centerline{
  	\begin{overpic}[height=6cm,tics=10]{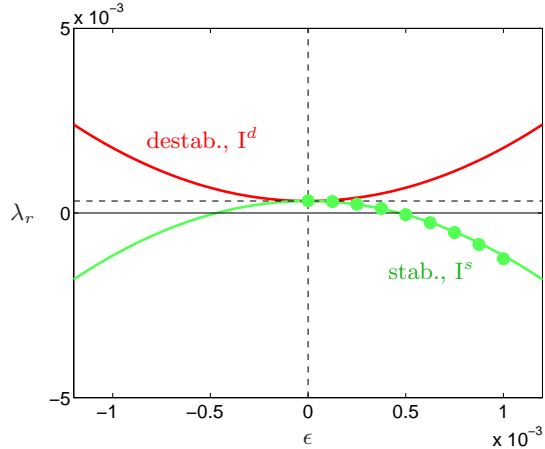}
	  	\put(24,57){\textcolor{red}{destab., I$^d$}} 
	  	\put(70,32){\textcolor{greenstab}{stab., I$^s$}} 
   	\end{overpic}  
  }    
   \caption{
   Effect of optimal 2D flow modification (spanwise-periodic, control wavenumber $\beta=1$)  on the growth rate of the leading eigenmode  $(\alpha_0,\beta_0)=(1,0)$.
Lines: 1D sensitivity prediction; symbols: full 2D stability calculations.
   }
\label{fig:POIS_compare1D}
\end{figure}

Figure~\ref{fig:POIS_compare1D} shows how the  growth rate of the leading eigenmode varies with  control amplitude, when most stabilising/destabilising flow modifications are optimised for control wavenumber $\beta=1$ and eigenmode $(\alpha_0,\beta_0)=(1,0)$.
In the stabilising case, the agreement between 1D sensitivity predictions (solid line) and full 2D calculations (symbols) is excellent. As expected, $\ev_r$ varies quadratically with $\epsilon$.

\subsection{Robustness}
\label{sec:POIS_robust}

A question of  interest is whether the most stabilising flow modification obtained for $\alpha_0=\alpha_{0,max}$ and $\beta_0=\beta_{0,max}$ is  stabilising at other wavenumbers too. In other words, is the optimisation robust?
We investigate this question by choosing flow modifications  optimised for the leading mode $(\alpha_0,\beta_0)=(1,0)$ and computing their effect on other modes $(\alpha_0,\beta_0) \neq (\alpha_{0,max},\beta_{0,max})$.

\begin{figure}
  \psfrag{lr}[t][]{$\ev_r$}
  \psfrag{li}[r][][1][-90]{$\ev_i$}  
  \centerline{
    \hspace{0.1cm}
   	\begin{overpic}[height=6.6cm,tics=10]{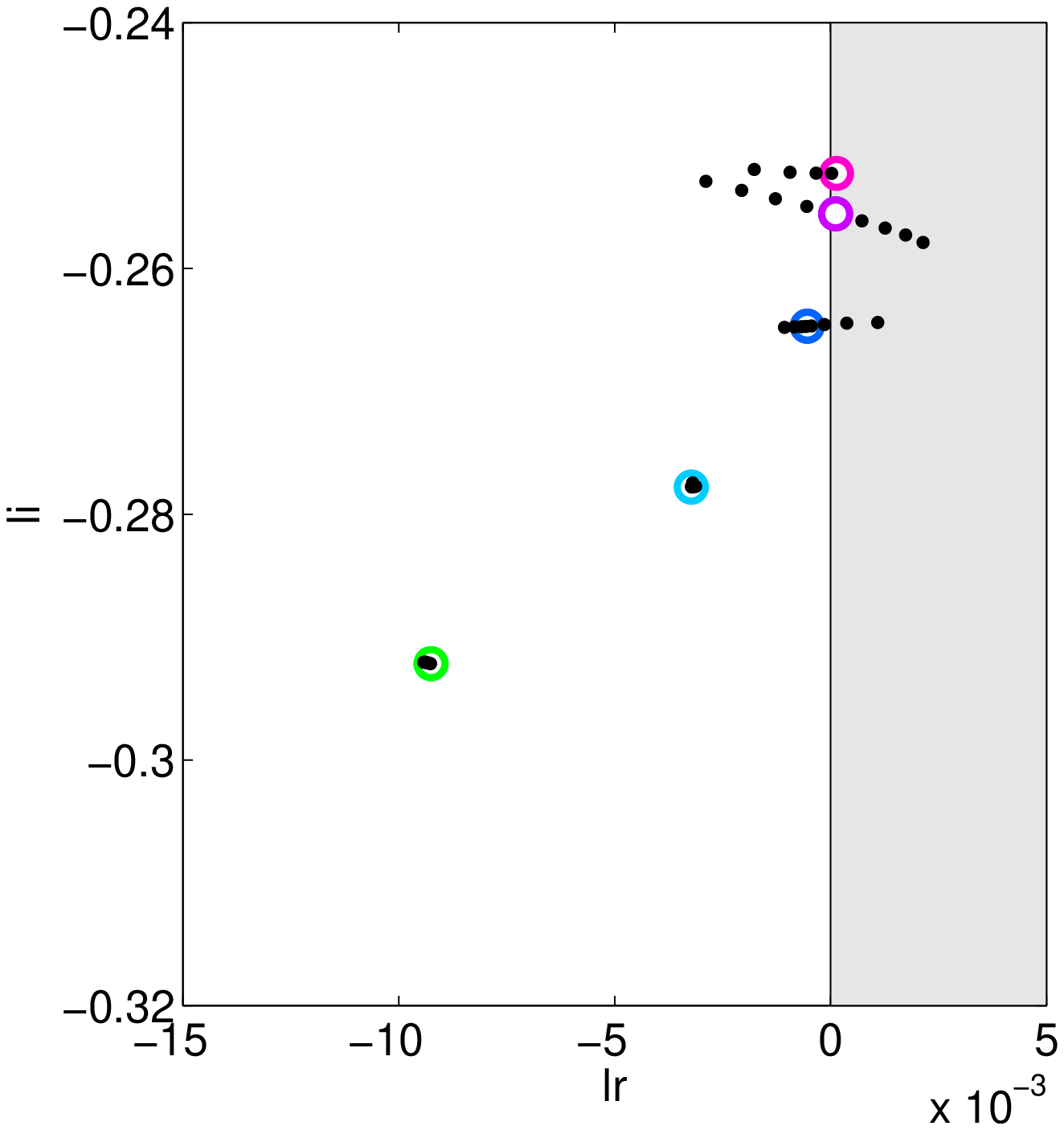}
		\put(-3,94){$(a)$}
   		\put(20,91){$\beta=0.4$}	
   		\put(77,84.5){$\beta_0=0$}	
 		\put(62,76.0){$\pm0.2$}	
  		\put(58,68.5){$\pm0.4$}
  		\put(48,54.5){$\pm0.6$}
  		\put(32,43.5){$\pm0.8$}	
   	\end{overpic}
    \hspace{0.6cm}
  	\begin{overpic}[height=6.6cm,tics=10]{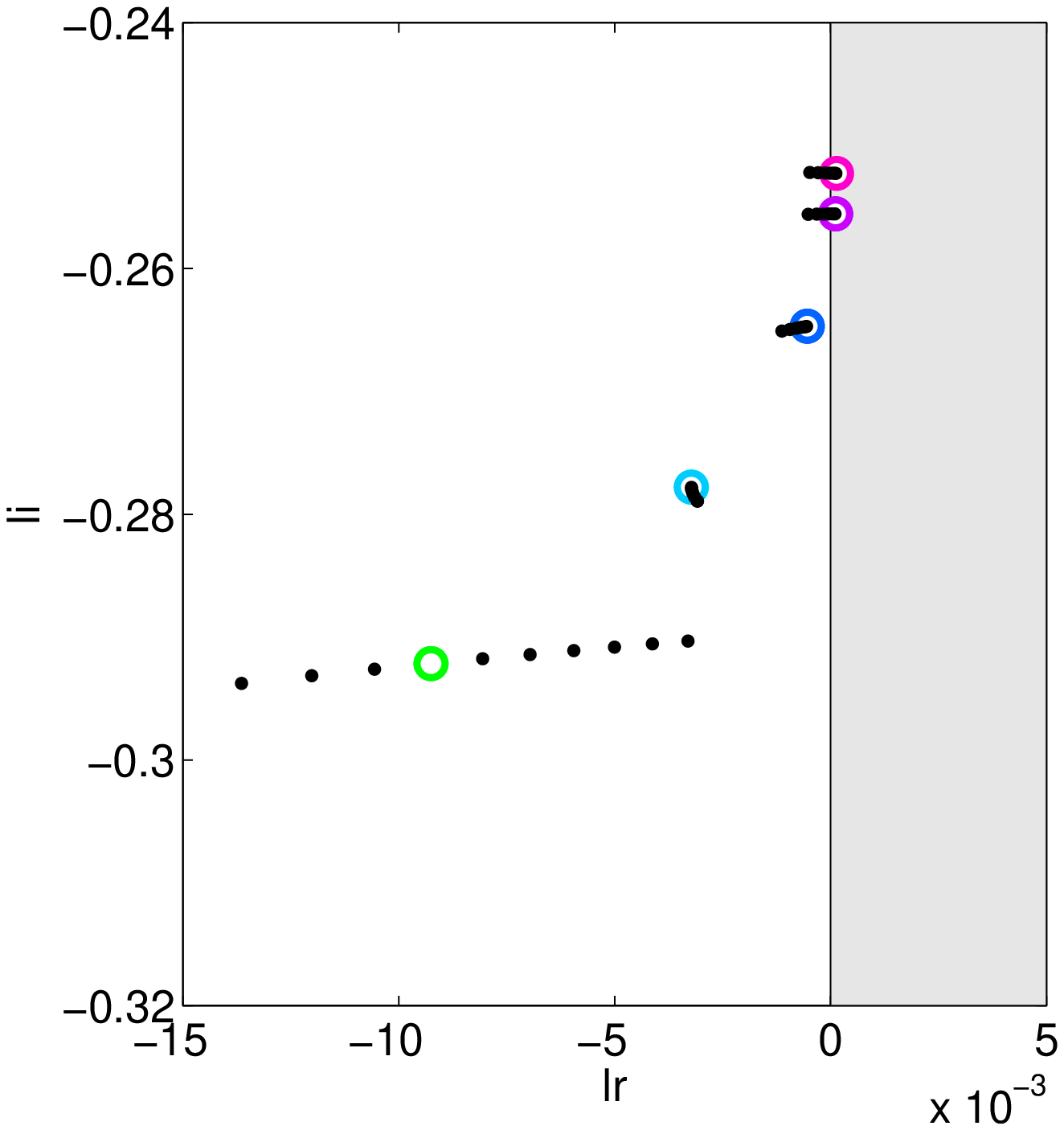}
    	\put(-3,94){$(b)$}
  		\put(20,91){$\beta=1.6$}
  		\put(77,84.5){$\beta_0=0$}	
		\put(62,76.0){$\pm0.2$}   	
 		\put(58,68.5){$\pm0.4$}   
  		\put(48,54.5){$\pm0.6$}
  		\put(32,43.5){$\pm0.8$}	
   	\end{overpic}
  }      
  \caption{
Effect of the most stabilising spanwise-periodic flow modification  $\epsilon U_1(y)\cos(\beta z)$ on leading 2D and 3D eigenvalues at $\alpha_0=1$, 
   with flow modification wavenumber  $(a)$~$\beta=0.4$ and $(b)$~$\beta=1.6$,
and $U_1$ optimised for the most unstable eigenmode $(\alpha_0,\beta_0)=(1,0)$.   
The 3D eigenpair with $\beta_0=\pm \beta/2$ undergoes a splitting, and one of the two modes becomes unstable. This issue is avoided by choosing  larger $\beta$ values, since the 3D eigenmodes affected by splitting are then very stable.
$(a)$ $\epsilon=(0.18, 0.35, 0.53, 0.71)\times10^{-3}$ , $(b)$ $\epsilon=(0.18\ldots, 0.71, 0.88, 1.06)\times10^{-3}$.
}
\label{fig:POIS_spectrum}
\end{figure}

Figure~\ref{fig:POIS_spectrum} shows eigenspectra when the base flow is modified with control wavenumbers $\beta=0.4$ and $\beta=1.6$.
The 2D mode $\beta_0=0$ is fully restabilised as predicted by sensitivity analysis.
Three-dimensional modes 
are stabilised or destabilised depending on $\beta$ and $\beta_0$.
A splitting of modes  $\beta_0=\pm\beta$ and $\beta_0=\pm\beta/2$ 
is systematically observed as a result of the flow modification and of a subharmonic resonance, 
and is consistent with Floquet analyses in other spatially periodic flows \citep{Herbert88,Hwang13}.
Therefore, the use of  small flow modification wavenumbers (fig.~\ref{fig:POIS_spectrum}$(a)$) yields a large stabilising effect on the leading 2D mode, as discussed in section \ref{sec:POIS_optimflowmodif}, but this also destabilises some slightly damped 3D eigenmodes. 
The choice of larger values of $\beta$ (fig.~\ref{fig:POIS_spectrum}$(b)$)  circumvents this issue  since the 3D modes that undergo splitting are more stable, and so remain at the amplitude $\epsilon$ needed to fully restabilise the 2D mode and thus the whole flow.

\begin{figure}
  \def\thisfigy{70} 
  \psfrag{alpha}[t][]{$\alpha_0$}
  \psfrag{lamr}[r][][1][-90]{$\ev_{r}$}  
  \centerline{
   	\begin{overpic}[width=6cm,tics=10]{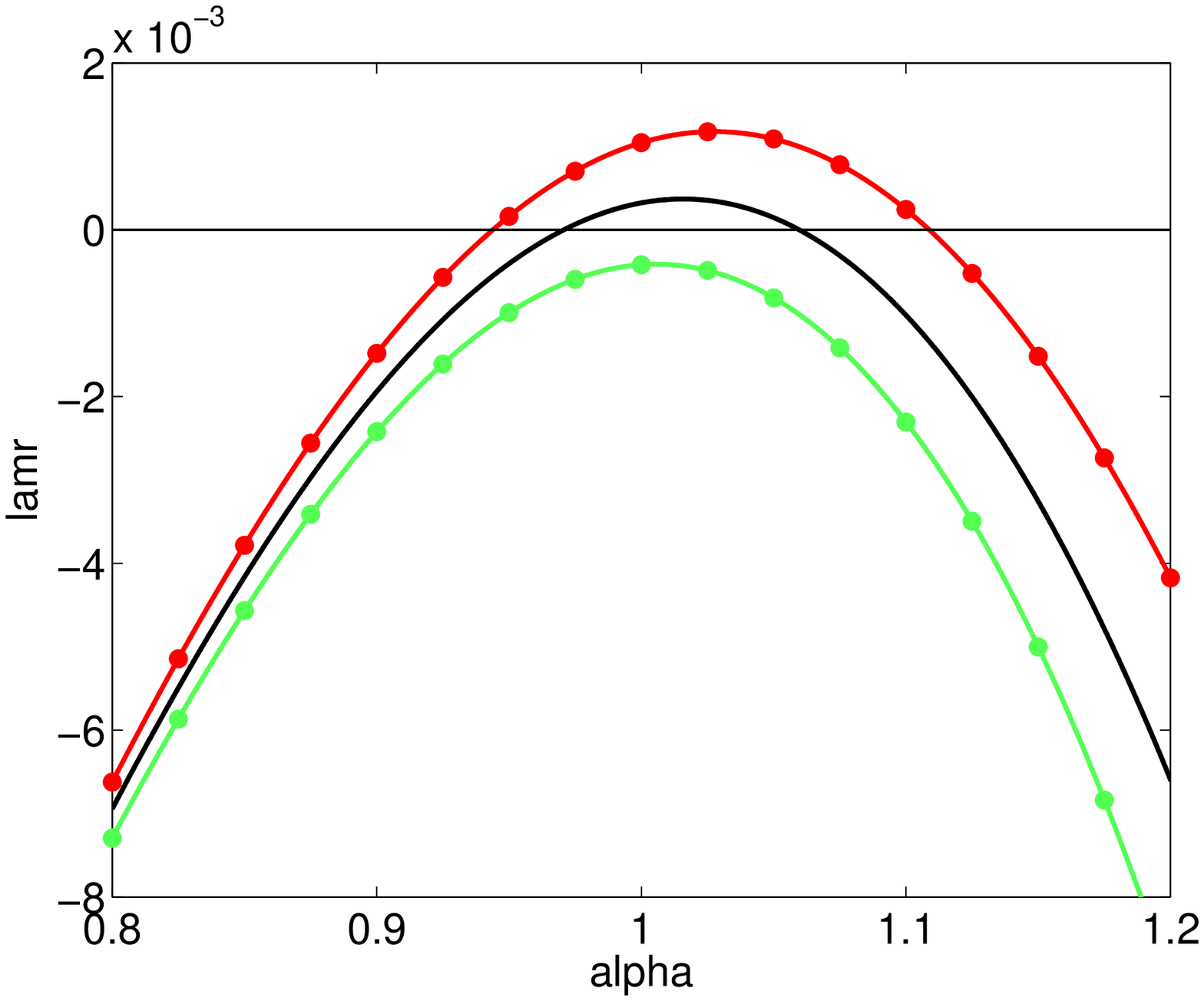}
	  	\put(15,\thisfigy){$\beta=1$}	
	  	\put(76,67){ \textcolor{red}      {I$^d$} }	
	  	\put(69,40){ \textcolor{greenstab}{I$^s$} }	
   	\end{overpic}
	\hspace{0.4cm}
   	\begin{overpic}[width=6cm,tics=10]{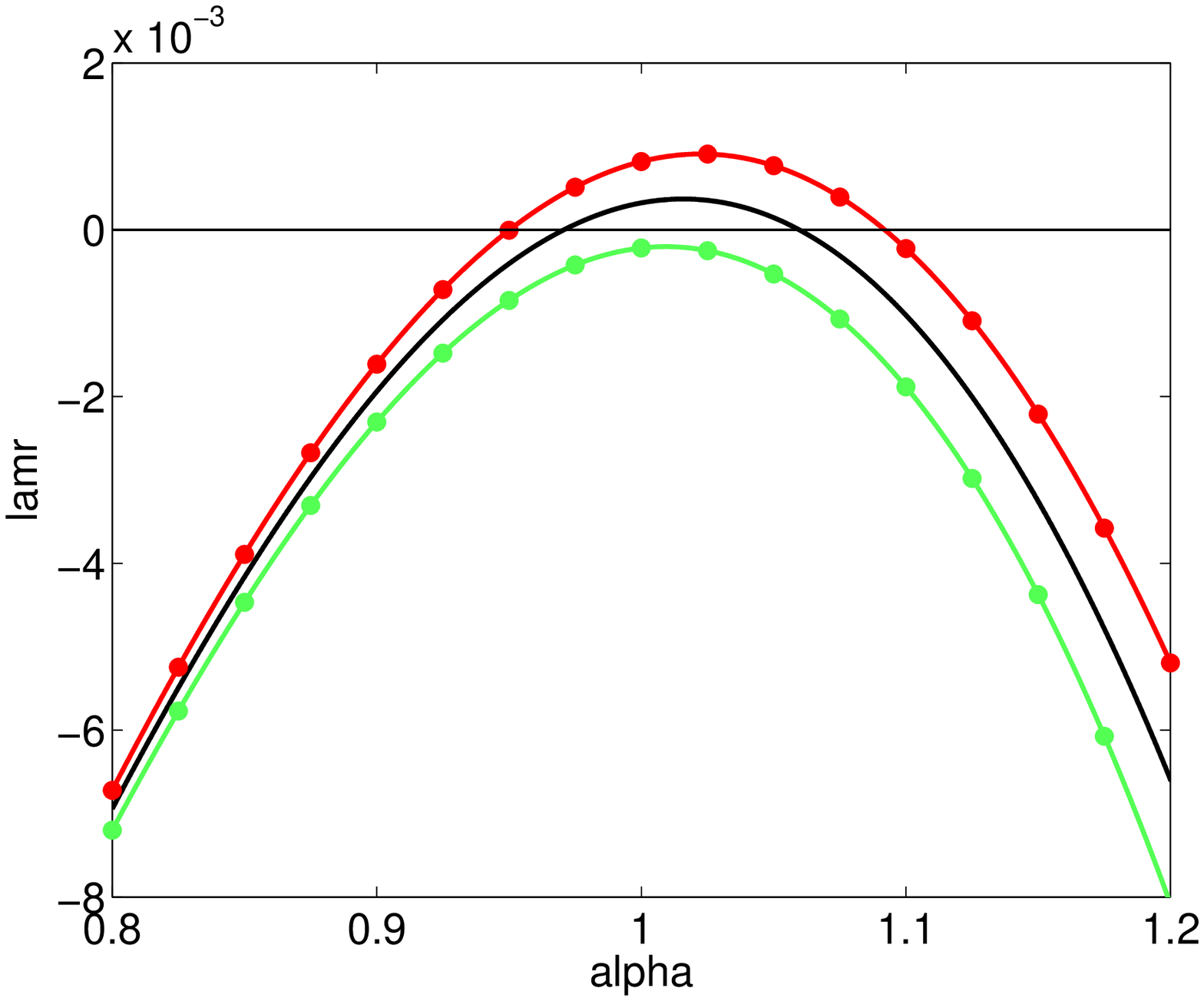}
		\put(15,\thisfigy){$\beta=3$}	
	  	\put(75,67){ \textcolor{red}      {I$^d$} }	
	  	\put(70,40){ \textcolor{greenstab}{I$^s$} }	
   	\end{overpic}
  }      
   \caption{Effect of optimal spanwise-periodic flow modification  $\epsilon U_1(y)\cos(\beta z)$ on the leading growth rate with $U_1$ optimised for $(\alpha_0,\beta_0)=(\alpha_{0,max},\beta_{0,max})=(1,0)$, at control wavenumber $\beta=1$ and 3.
   Branch I$^d$: most destabilising;  branch I$^s$: most stabilising.
   $\epsilon=0.001$.}
\label{fig:POIS_robust}
\end{figure}

Next, we look at robustness in $\alpha_0$. 
Figure~\ref{fig:POIS_robust} shows that whatever the choice of spanwise wavenumber $\beta$, flow modifications optimised to stabilise (destabilise) the leading mode at $\alpha_{0,max}$
have a stabilising (destabilising) effect at  all other values of $\alpha_0$ too.

\bigskip
In this section we computed the optimal spanwise-periodic flow modifications yielding the largest second-order variation in growth rate, and showed that is was possible to fully restabilise the plane channel flow with small-amplitude modifications.
In the next section, we turn to  a more strongly unstable flow and investigate to what extent this strategy remains effective and robust.

\section{Results: the parallel mixing layer}
\label{sec:mixinglayer}

We now focus on  the hyperbolic-tangent mixing layer  $U_0(y)=1+R\tanh(y)$ \citep{Michalke64} with $R=1$, at $\Rey=100$.
Unlike the plane channel flow of section~\ref{sec:planechannel}, which is unstable only in the neighbourhood of  $\alpha_{0,max}=1$ with  viscous eigenmodes characterised by weak growth rates,   the mixing layer is unstable  in the whole band of wavenumbers $0\leq\alpha_0\leq1$, and its inviscid eigenmodes exhibit much stronger growth rates.
 We use the same 1D spectral method as in section~\ref{sec:planechannel} on the domain $y\in[-5;5]$ with  homogeneous Dirichlet boundary conditions on velocity components.
The largest growth rate is obtained for the streamwise wavenumber 
$\alpha_0=\alpha_{0,max}=0.45$, where the leading eigenvalue is  $\ev_0=0.1676 - 0.4500i$, in  agreement with existing results \citep{Betchov63, Michalke64, Villermaux98}.

\subsection{Optimal flow modifications}
\label{sec:TANH_optimflowmodif}

\begin{figure}
  \psfrag{beta}[t][]{$\beta$}
  \psfrag{alpha}[t][]{$\alpha_0$}
  \psfrag{lam2}   [][]{} 
  \psfrag{lam2max}[b][]{}
  \psfrag{lam2ds}[b][]{} 	
  \psfrag{betamax}[b][]{} 	
  \vspace{0.5cm}
  \centerline{
  	\hspace{-0.35cm}
  	\begin{overpic}[height=5.46cm,tics=10]{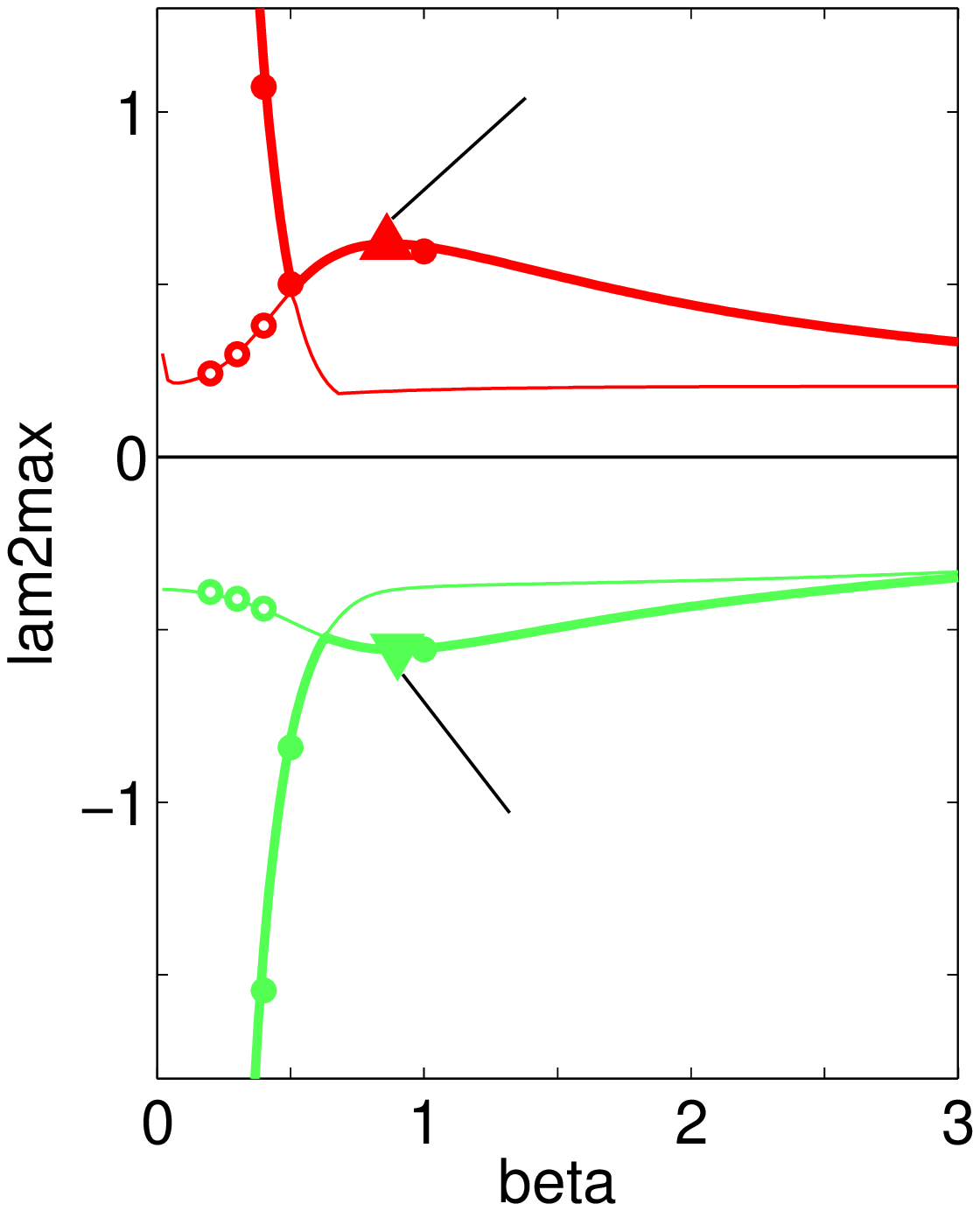}
	  	\put(10,105.5){$(a)$ $\min \ev_{2r}$, $\max \ev_{2r}$}
		\put(25,91){\textcolor{red}{II$^d$}} 
  		\put(68,75){\textcolor{red}{I$^d$}}
  		\put(42,89){ $(\beta^d, \ev_{2r}^d)$}
  		\put(41,32){ $(\beta^s, \ev_{2r}^s)$} 		
  		\put(25,15){\textcolor{greenstab}{II$^s$}}
  		\put(69,45){\textcolor{greenstab}{I$^s$}}
   	\end{overpic}
   	\hspace{-0.1cm}
  	\begin{overpic}[height=5.59cm,tics=10]{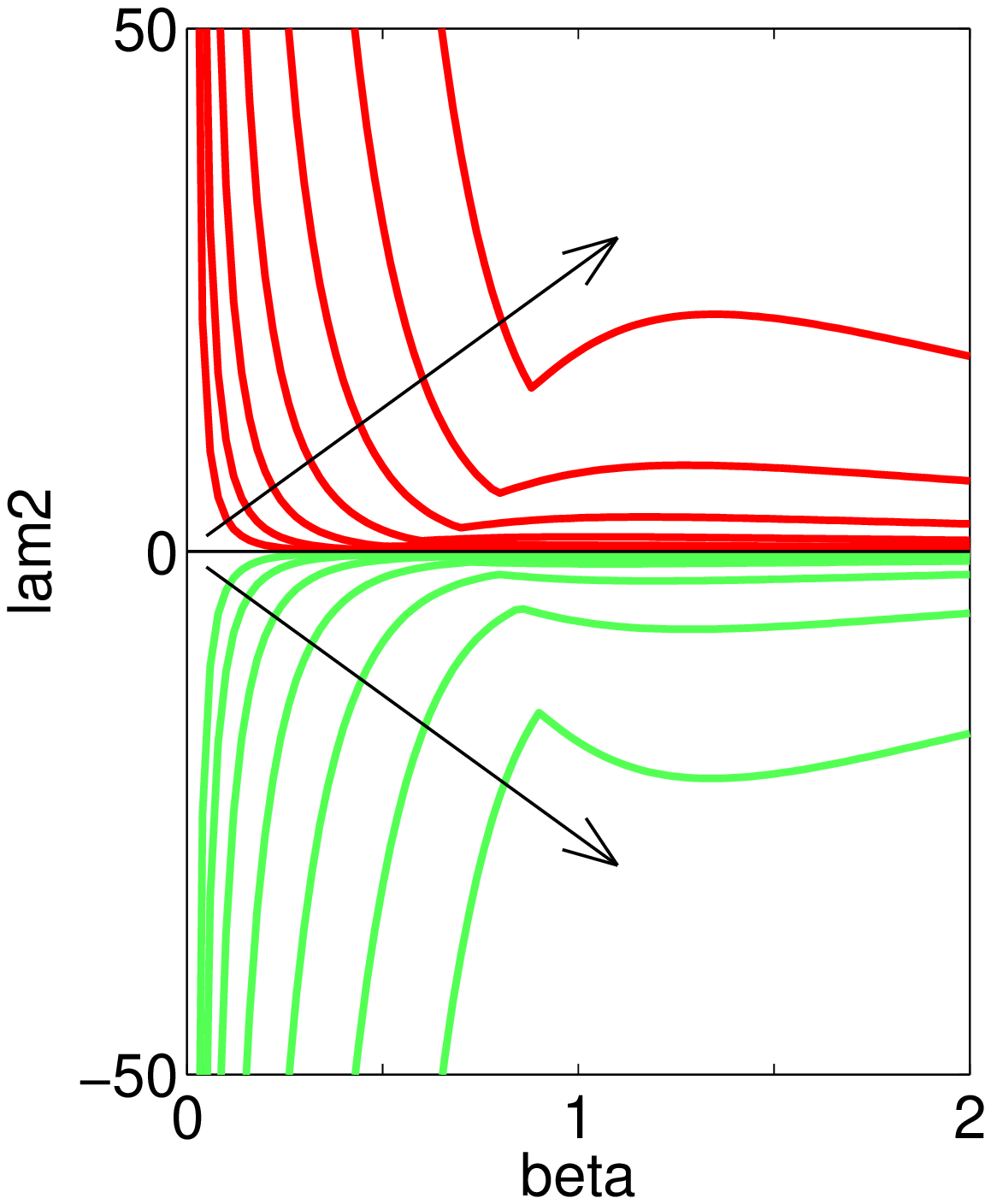}
	  	\put(9,103){$(b)$ $\min \ev_{2r}$, $\max \ev_{2r}$} 
	  	\put(41,82){\footnotesize $\alpha_0=0.3\ldots0.9$}  
   	\end{overpic}
   	\hspace{0.05cm}
   	\begin{overpic}[height=5.68cm,tics=10]{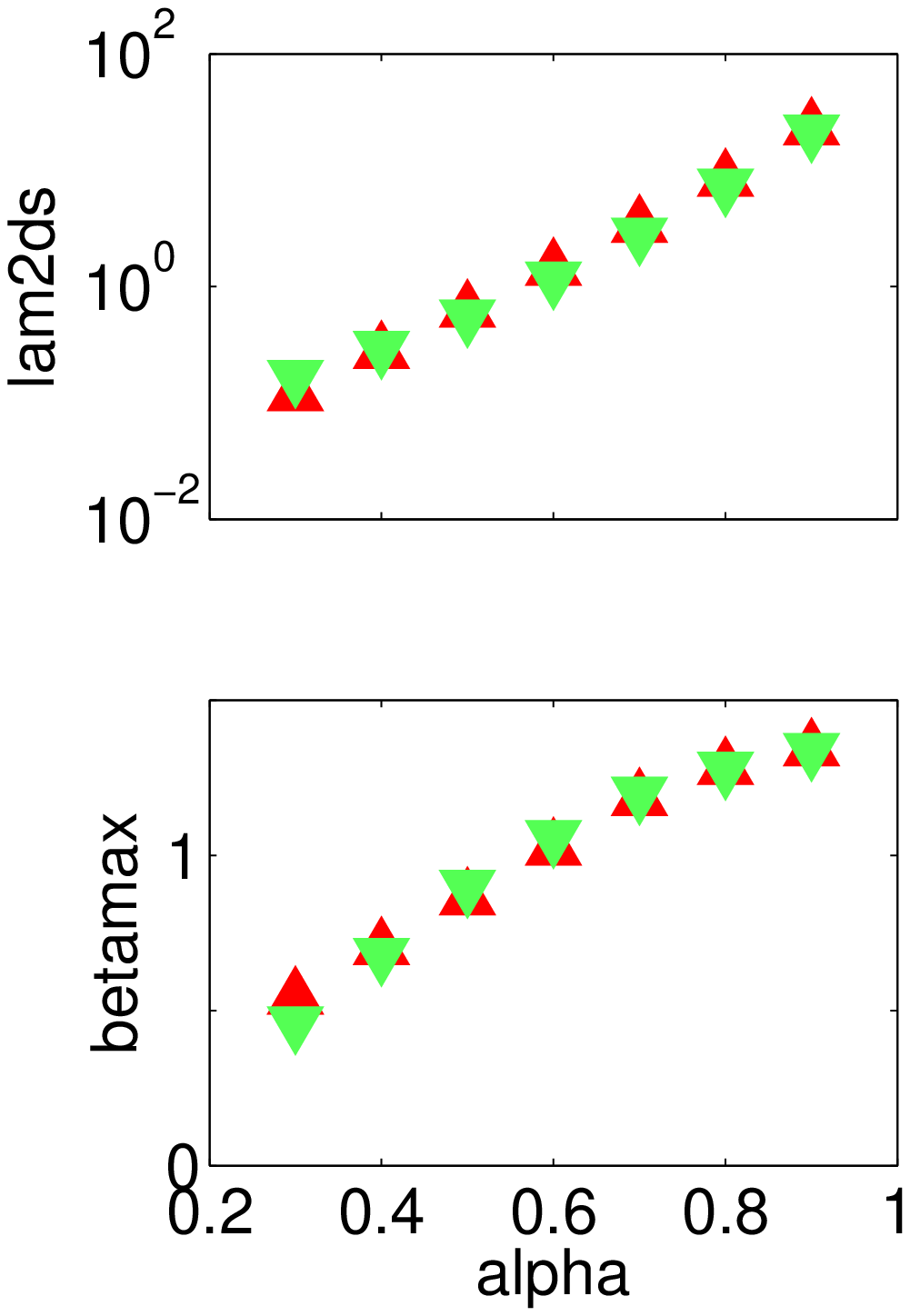}
 	  	\put(7,101.5){$(c)$}
   		\put(20,88){$\ev_{2r}^d$, $\ev_{2r}^s$}	
   		\put(20,39){$\beta^{d}$, $\beta^{s}$}
   	\end{overpic}
  }   
  \vspace{-0.01cm} 
  \caption{Upper and lower bounds on $\ev_2$, i.e. maximal  destabilising (dark, red online) and stabilising (light, green online)  effect on the leading growth rate, as predicted by sensitivity analysis. 
   $(a)$~$\alpha_0=0.5$. Crossing of branches I and II, optimal at large and small $\beta$ respectively. 
   Circles are calculations for the full stability problem.
   $(b)$~Upper and lower bounds for $\alpha_0=0.3 \ldots 0.9$.
   $(c)$~Variation with  $\alpha_0$ of the local maxima and minima shown as triangles in $(a)$, and corresponding spanwise wavenumbers. 
 }
\label{fig:TANH_lam_vs_beta}
\end{figure}

We compute the most destabilising and most stabilising  spanwise-periodic flow modifications according to (\ref{eq:mostdestab}).
Figure~\ref{fig:TANH_lam_vs_beta}$(a)$ shows the largest positive and negative second-order eigenvalue variations at $\alpha_0=0.5$ as a function of the control spanwise wavenumber.
Again, these curves provide bounds for the largest possible destabilisation and stabilisation. 
Each of these two curves has a local extremum, $\ev_{2r}^{d}$ and $\ev_{2r}^{s}$ respectively, close to $\beta^{d}=\beta^{s}=0.8$.
As shown in figure~\ref{fig:TANH_lam_vs_beta}$(b)$,
for a fixed choice of $\beta$ the upper and lower bounds strongly increase  with  $\alpha_0$, indicating that optimal flow modifications have a stronger effect on the leading eigenmode at smaller streamwise wavelengths.
Figure~\ref{fig:TANH_lam_vs_beta}$(c)$ shows that the local extrema $\ev_{2r}^{d}$ and $\ev_{2r}^{s}$ increase exponentially, while the corresponding values of $\beta$ become slightly larger but remain of order $\sim$1.
For smaller values of $\beta$, the upper and lower bounds diverge like $\beta^{-2}$, like in the plane channel flow.

\begin{figure}
  \psfrag{U1}[t][]{$U_1$}
  \psfrag{U0+U1d}[t][]{ }
  \psfrag{U0+U1s}[t][]{$U_0\pm\epsilon U_1$}
  \vspace{0.2cm}
  \centerline{
  	\begin{overpic}[height=3.5cm,tics=10]{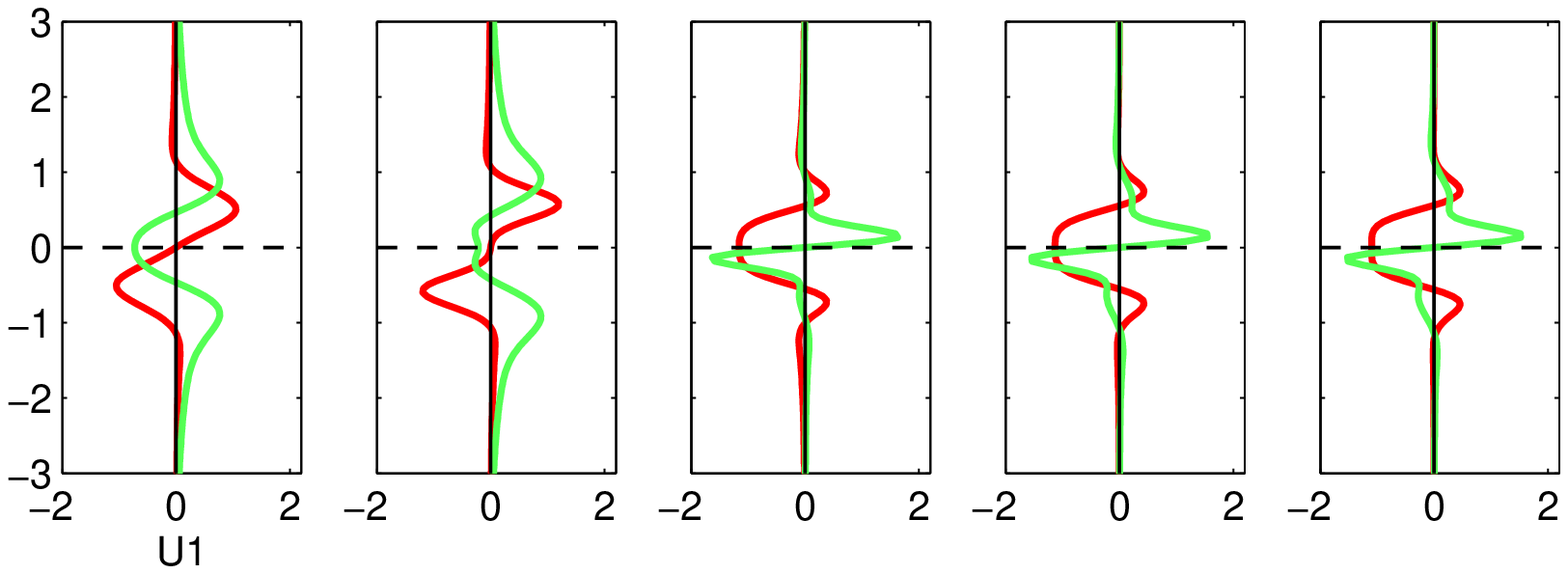}
   		\put(-11.5,34){$(a)$}
   		\put(-5,20.5){$y$}
  		\put( 6,37){$\beta=0.2$}
  		\put(26,37){$\beta=0.4$}
  		\put(46,37){$\beta=0.6$}
  		\put(66,37){$\beta=0.8$}
  		\put(87,37){$\beta=1$}
   	\end{overpic}
  }    
  \vspace{0.2cm}
  \centerline{
  	\begin{overpic}[height=3.5cm,tics=10]{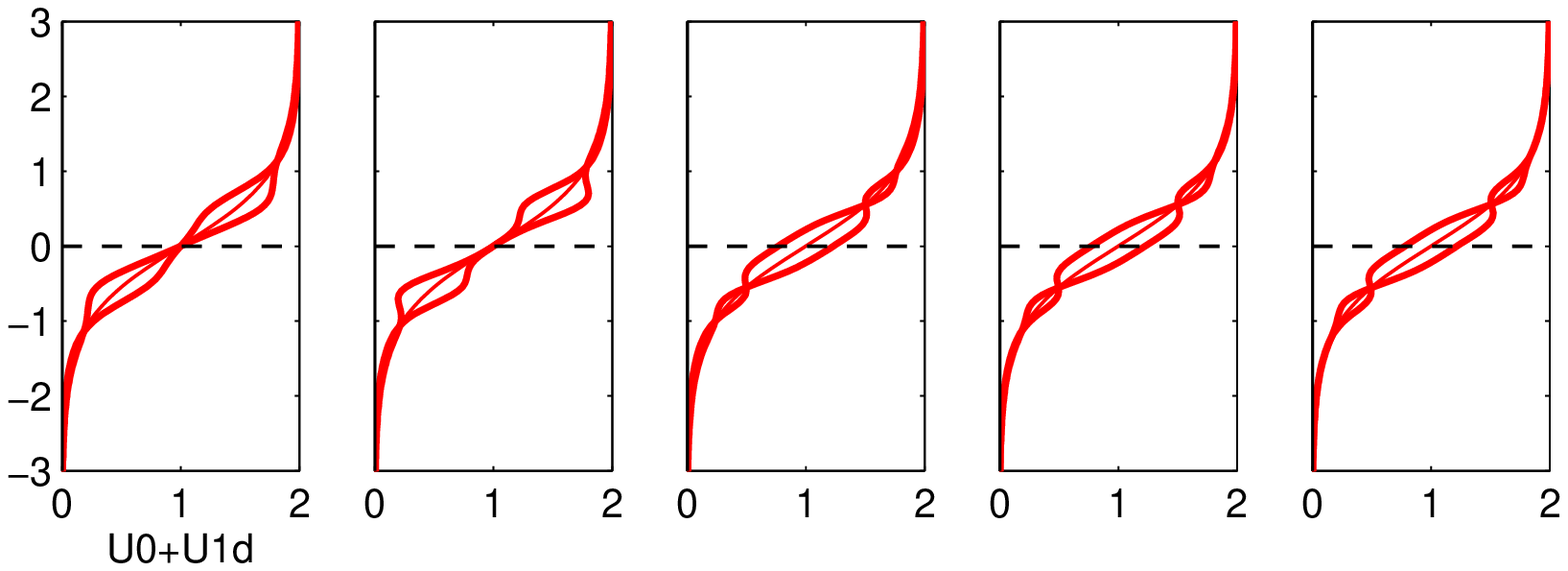}
  		\put(-11,33.5){$(b)$}
  		\put(-5,20.5){$y$}
  		\put(102,20.5){\textcolor{red}{dest.}}
   	\end{overpic}  
  }   
  \vspace{-0.5cm}
  \centerline{
  	\begin{overpic}[height=3.5cm,tics=10]{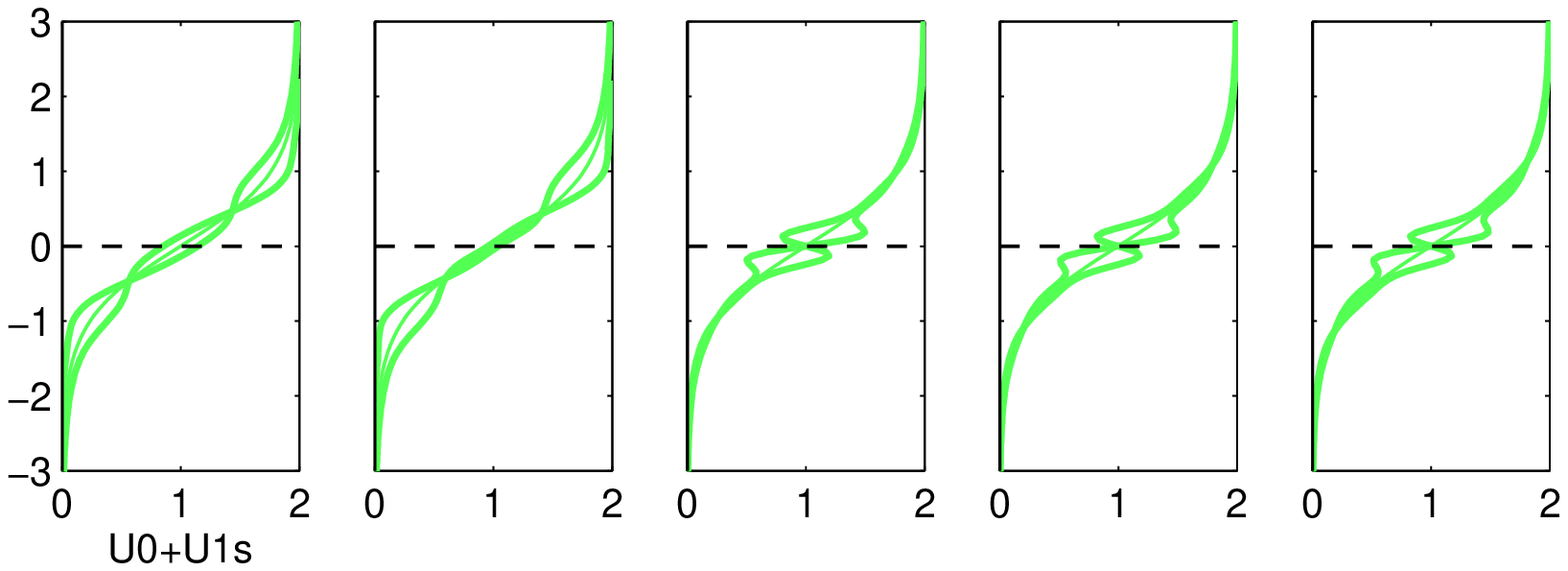}
  		\put(-5,20.5){$y$}
 		\put(102,20.5){\textcolor{greenstab}{stab.}}
   	\end{overpic}  
  }
   \caption{$(a)$ Most destabilising (dark, red online) and stabilising (light, green online) for $\alpha_0=\alpha_{0,max}=0.45$, $\beta=0.2$, 0.4, 0.6, 0.8, 1.
   $(b)$ Total modified flow $U_0(y)+\epsilon U_1(y)\cos(\beta z)$ at  $z=0$ and $z=\pi/\beta$ for $\epsilon=0.2$.}
\label{fig:TANH_U0U1branchI}
\end{figure}

The upper and lower bounds in figure \ref{fig:TANH_lam_vs_beta}$(a)$ are actually made of two branches which intersect and correspond to different families of flow modifications $U_1$.
We call branch I the optimal family at large $\beta$ (corresponding to the local extrema at $\beta^d$, $\beta^s$) and branch II the optimal family  at small $\beta$ (diverging as $\beta \rightarrow 0$).
Optimal modifications  are shown in figure~\ref{fig:TANH_U0U1branchI} 
for $\alpha_0=\alpha_{0,max}=0.45$.
At small spanwise wavelength $\beta < 0.5$, the most destabilising $U_1$ is antisymmetric and the most stabilising $U_1$ is symmetric;
at larger spanwise wavelength $\beta > 0.5$,  symmetry properties are exchanged as branches I and II cross.

\subsection{Validation}
\label{sec:TANH_valid}

Similarly to section \ref{sec:POIS_valid}, we compare the effect of optimal spanwise-periodic flow modifications of unit 1D norm predicted by sensitivity analysis and obtained from full 2D calculations 
with the same numerical method as in sections~\ref{sec:POIS_valid}-\ref{sec:POIS_robust} on $y\in[-5;5]$, $z\in[0; 2\cdot 2\pi/\beta]$.

\begin{figure}
  \psfrag{k=0.45}[t][]{}
  \psfrag{U}[t][]{$U_1$}
  \psfrag{eps}[t][]{$\epsilon$}
  \psfrag{y}[][][1][-90]{$y$}
  \psfrag{real(lambda0)}[][][1][-90]{$\ev_{r}\quad$}
  \centerline{ 	
  	\begin{overpic}[height=6cm,tics=10]{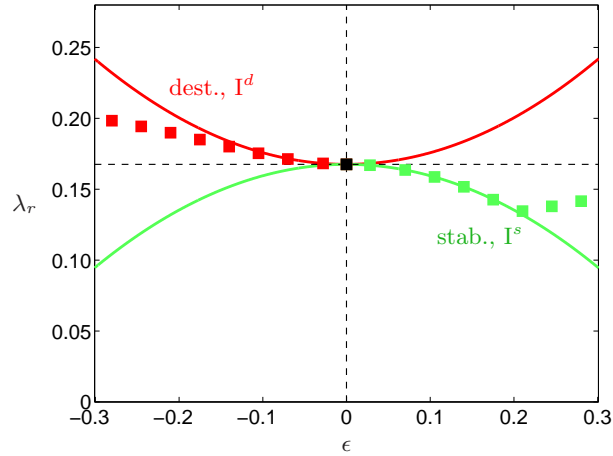}
	  	\put(25,60){\textcolor{red}      {dest., I$^d$}}
	  	\put(70,35){\textcolor{greenstab}{stab., I$^s$}}	
   	\end{overpic}  
  }    
   \caption{
    Effect of optimal 2D (spanwise-periodic) at $\beta=0.8$ on the leading growth rate at $\alpha_0=\alpha_{0,max}=0.45$. Lines: sensitivity prediction; symbols: full stability calculations.
   }
\label{fig:TANH_compare1D}
\end{figure}

The variation of the leading growth rate $(\alpha_0,\beta_0)=(0.45,0)$ for control wavenumber  $\beta=0.8$ (branch I) is  illustrated in figure~\ref{fig:TANH_compare1D}. 
The agreement is good at small amplitudes $|\epsilon| \lesssim  0.2$.  
At larger amplitudes the actual growth rate departs from the expected quadratic dependence, due to non-linear effects and to other modes becoming unstable. We have not investigated  the splitting described in section~\ref{sec:POIS_robust} systematically.

\subsection{Robustness}
\label{sec:TANH_robust}

\begin{figure}
  \psfrag{alpha}[t][]{$\alpha_0$}
  \psfrag{lamr}[][]{$\ev_{r}$}  
  \hspace{0.2cm}
  \centerline{
  	\begin{overpic}[width=13cm,tics=10]{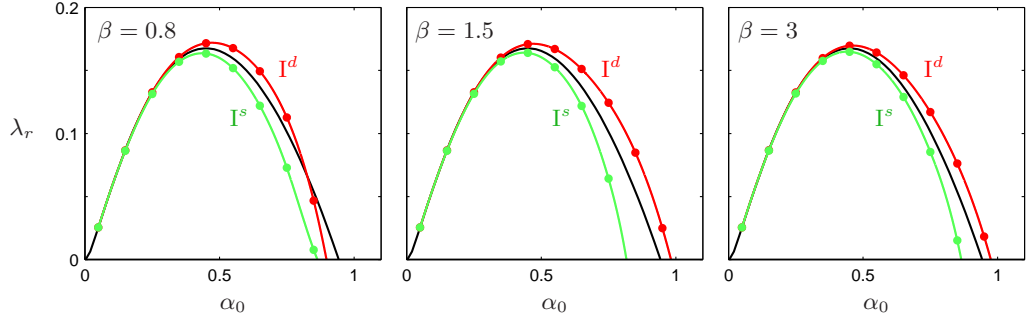}
	  	\put(4.5,28){$\beta=0.8$}	
	  	\put(23,24){\textcolor{red}{I$^d$}}	
	  	\put(18,19){\textcolor{greenstab}{I$^s$}}
	  	\put(37,28){$\beta=1.5$}	
	  	\put(56,24){\textcolor{red}{I$^d$}}	
	  	\put(51,19){\textcolor{greenstab}{I$^s$}}	  	
	  	\put(70,28){$\beta=3$}	
	  	\put(89,24){\textcolor{red}{I$^d$}}	
	  	\put(84,19){\textcolor{greenstab}{I$^s$}}	
	  	\put(-4.5,18){$\ev_r$}
   	\end{overpic}
  }      
   \caption{Effect of optimal spanwise-periodic flow modification  $\epsilon U_1(y)\cos(\beta z)$ on the leading growth rate with $U_1$ optimised for $\alpha_0=\alpha_{0,max}=0.45$ and  $\beta=0.8$, 1.5, 3.
   Branch I$^d$: most destabilising;  branch I$^s$: most stabilising.
   $\epsilon=0.1$.}
\label{fig:TANH_robust}
\end{figure}

Since the mixing layer is unstable over a wide range of streamwise wavenumbers, it is important to assess the robustness of the flow modifications designed for the most unstable mode $\alpha_0=\alpha_{0,max}$.
We investigate this point by computing  the variation in the leading growth rate predicted by (\ref{eq:ev23}) when the flow is modified with $\epsilon U_1(y)\cos(\beta z)$, where $\epsilon=0.1$ and  $U_1$  is chosen in branch I for $\alpha_0=\alpha_{0,max}=0.45$. 
Figure~\ref{fig:TANH_robust} shows that whatever the choice of spanwise wavenumber $\beta$, flow modifications optimised to stabilise (destabilise) the leading mode at 
$\alpha_{0,max}$
have a stabilising (destabilising) effect at almost all other values of $\alpha_0$ too.
This effect is negligible at small $\alpha_0$, and larger at larger $\alpha_0$. 
This is due to the dispersion relation at small $\alpha_0$ being independent of the details of the velocity profile: 
in particular, according to the Kelvin-Helmholtz dispersion relation 
 $\ev_r=\alpha_0 \Delta U/2$ pertaining to the vorticity sheet model, the growth rate is  only determined by the velocity difference $\Delta U$ between the two streams.
In contrast, the maximal growth rate and the cut-off wavenumber are influenced by other characteristics of the velocity profile (e.g. thickness and shear).
Here, since $U_1(y)$ vanishes far from $y=0$,  the velocity difference $\Delta U=2R$ remains constant, and so does the growth rate at small $\alpha_0$.

\bigskip
In this section we showed that strongly unstable eigenmodes in the mixing layer were made more stable by adding  optimal 2D spanwise-periodic modifications to the base flow. 
Modifications designed for the most unstable  streamwise wavenumber $\alpha_{0,max}$ have a stabilising effect on other unstable wavenumbers $\alpha_0$. A wide range of modification wavenumbers $\beta$ were found to be effective.
In the next section, we investigate how the impact of 2D spanwise-periodic modifications can be increased by taking advantage of non-normal transient growth, and achieving large modifications from initially small perturbations.

\section{Leveraging transient growth to increase stabilisation}
\label{sec:combined}

Non-normal mechanisms can lead to substantial transient growth in many flows \citep{Butler92,Trefethen93,Schmid01}. 
This phenomenon is thought to make possible subcritical transition to turbulence in linearly stable flows, like the flow in a circular pipe,
since small-amplitude perturbations can undergo large (linear) amplification and eventually trigger (non-linear) destabilisation.
This potential for large amplification  has also been exploited as a control strategy: it has been observed that streamwise vortices amplified into streamwise streaks through the lift-up mechanism are able to stabilise boundary layers  \citep{Cossu02,Fra05} and wakes \citep{DelGuercio2014}.
It remains unclear, however, 
whether optimally amplified perturbations always have a stabilising effect, and 
whether there exist other perturbations that undergo a smaller amplification but eventually yield a larger stabilisation.

In this section we revisit this control strategy in terms of optimal flow modification.
\textcolor{black}{
We consider the stabilising effect of spanwise-periodic perturbations undergoing transient amplification. 
This is similar in spirit to the study of 
\cite{DelGuercio2014}, but instead of computing 
\textit{a posteriori} the stabilising effect of optimally amplified perturbations, we rather optimise 
\textit{simultaneously} transient growth and stabilising effect at the time of maximal amplification, and we thus determine  
\textit{combined}  optimal perturbations.
It should be noted that we do not aim at complete restabilisation but rather wish to determine what kind of structure yields the largest overall stabilising effect.
The approach relies on an extension of the optimisation method presented earlier  (section \ref{sec:optimal}).
}

While the stabilisation/destabilisation of streamwise streaks has been shown in the literature to depend on their amplitude, we consider the effect of arbitrary streamwise-invariant spanwise-periodic structures under the strong hypothesis of their linear transient evolution. This is an important limitation of the present analysis. 
As the streaks evolve non-linearly, two effects may contribute to mitigate the present results: 
first the amplitude may saturate \citep{Cossu04} which would yield a quantitative difference;
second and more importantly, non-linear saturated streaks remain spanwise periodic but they lose their pure sinusoidal character in $z$, as higher harmonics are generated.

Another important limitation of the proposed approach is to consider, following  \cite{Cossu04} and
 \cite{Reddy98}, that a separation of time scales applies between the fast scale of the exponential instability of the nominal base flow and the rather slow scale of  transient growth mechanisms.

Therefore, the proposed combined optimisation is not expected to yield quantitatively accurate results, but rather constitutes a means to investigate the mechanisms at hand in streaky flows. 
As explained later on, it will allow us to interpret streamwise streaks as particularly efficient structures that  both benefit from transient growth and effectively stabilise the flow.

\subsection{Combined optimisation}
\label{sec:combinedmethod}

We first leave stabilisation aside  and recall the concept of  transient growth.
Denoting energy amplification from 0 to $t$ as $G(t) = ||\uu(t)||^2 / ||\uu(0)||^2$, 
and $\evol$ the linear operator that propagates perturbations in time according to (\ref{eq:LNS}) such that $\uu(t)=\evol(t)\uu(0)$,
the optimal transient growth can be computed at every time $t$ as
\begin{subeqnarray}
G_{opt}(t) 
= \max_{||\uu(0)||=1}  G(t) 
&=&  \max_{||\uu(0)||=1}   \ps{\uu(t)}{\uu(t)}  
\\
&=&  \max_{||\uu(0)||=1}   \ps{\evol(t)\uu(0)}{\evol(t)\uu(0)}  
\\
&=& \ev_{max} \left\{ \evol^\dag \evol \right\},
\end{subeqnarray}
and the corresponding optimal initial perturbation is the associated eigenvector $\uu_{opt}(0)$. 
The largest value of optimal transient growth over all times is reached for some time $T$: $G_{opt}(T) = \max_{t} G_{opt}(t)$.

\begin{figure}
   \psfrag{z}[t][]{$\beta z$}
   \psfrag{t}[t][]{$t$}
   \psfrag{y}[][][1][-90]{$y\quad$}
   \psfrag{G}[][][1][-90]{$G_{opt}\qquad$}
   \vspace{1cm}
   \centerline{	
  	\begin{overpic}[height=5cm,tics=10]{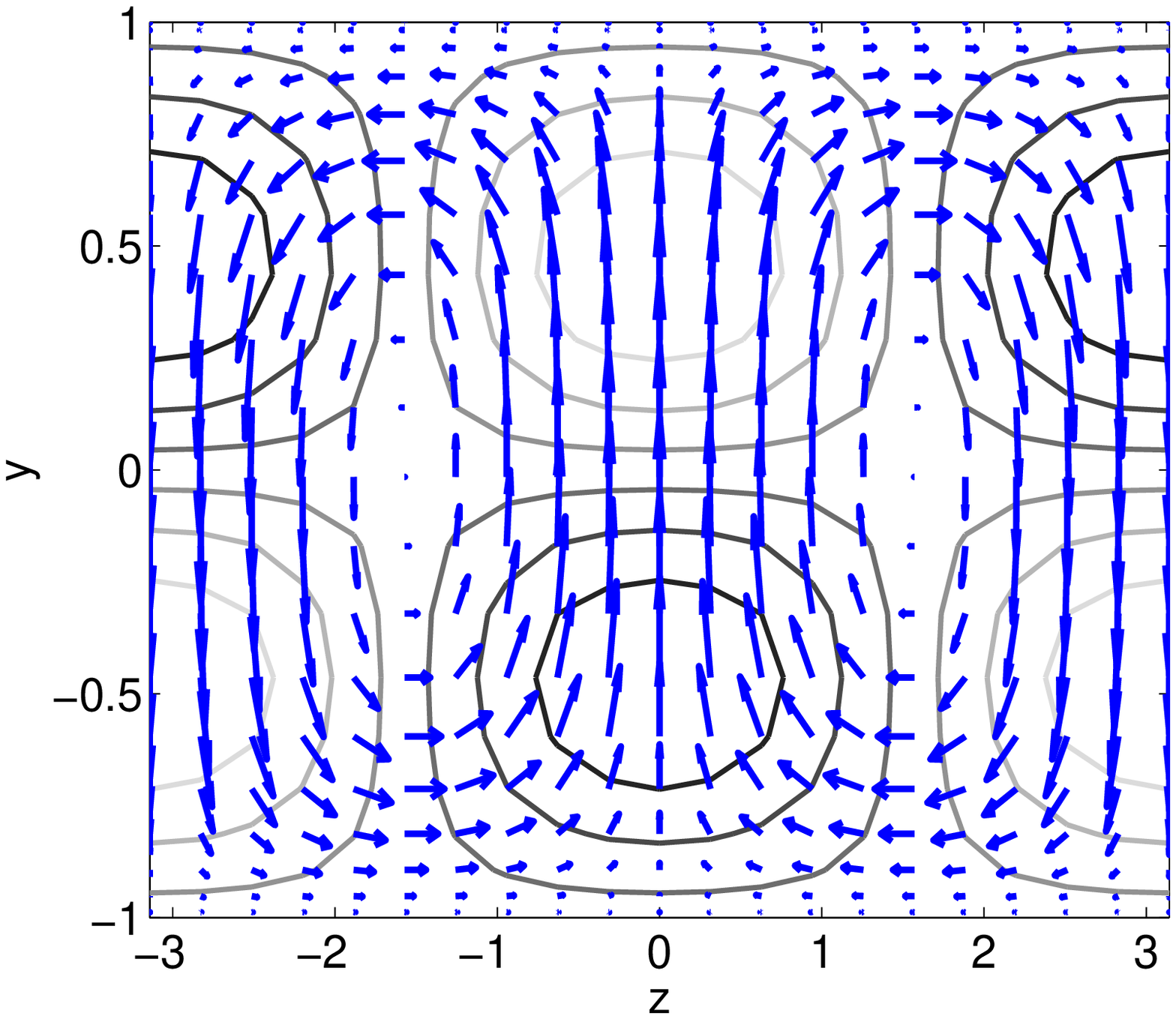}
  		\put(-3,80.5){$(a)$}
  		\put(40,89){Plane channel}
   	\end{overpic}  
   	\hspace{0.9cm}
   	\begin{overpic}[height=5cm,tics=10]{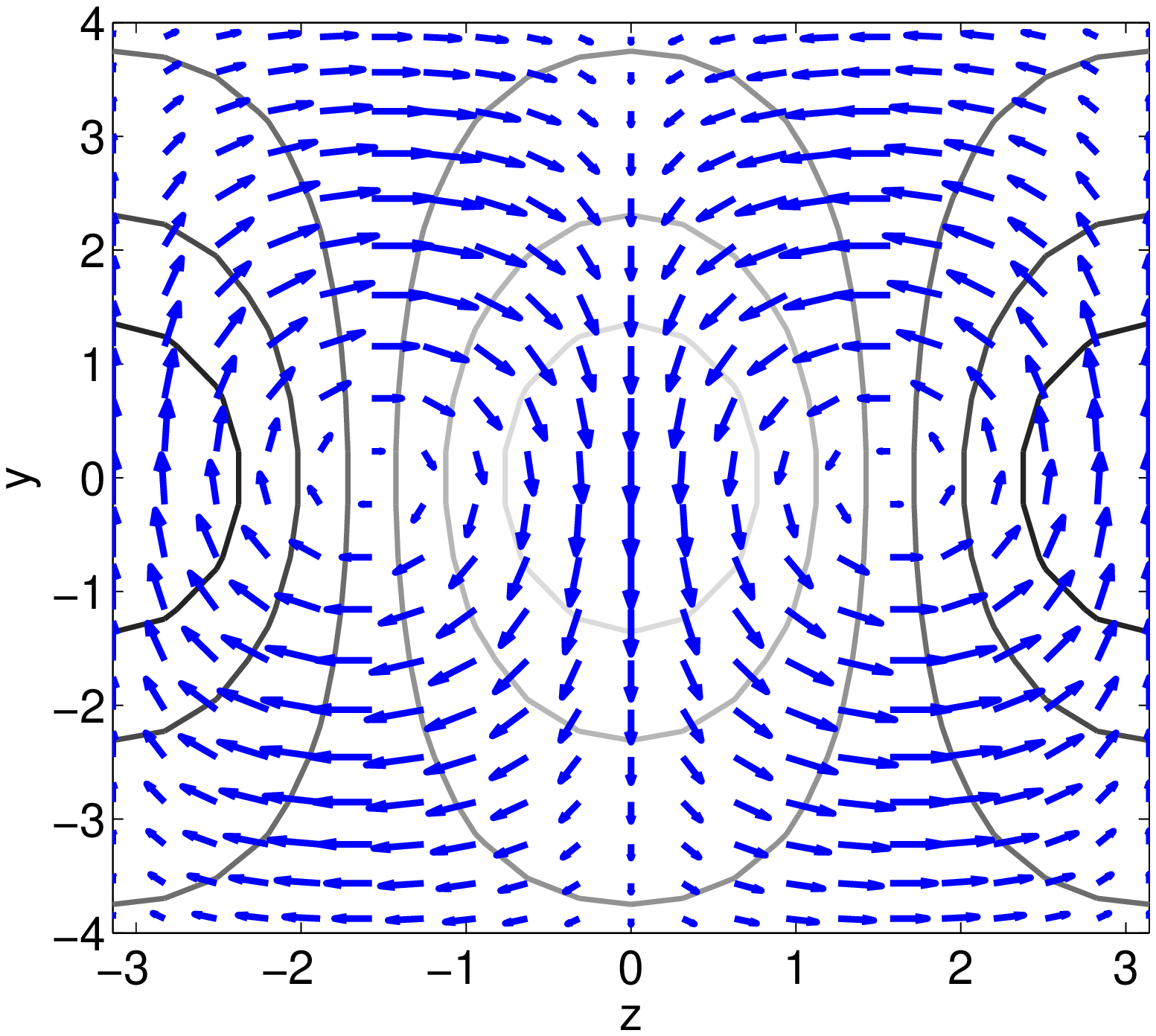}
  		\put(-3.5,82.5){$(b)$}   		
  		\put(40,91.5){Mixing layer}
   	\end{overpic} 
   } 
   \vspace{0.3cm}  
   \centerline{	
  	\begin{overpic}[height=5cm,tics=10]{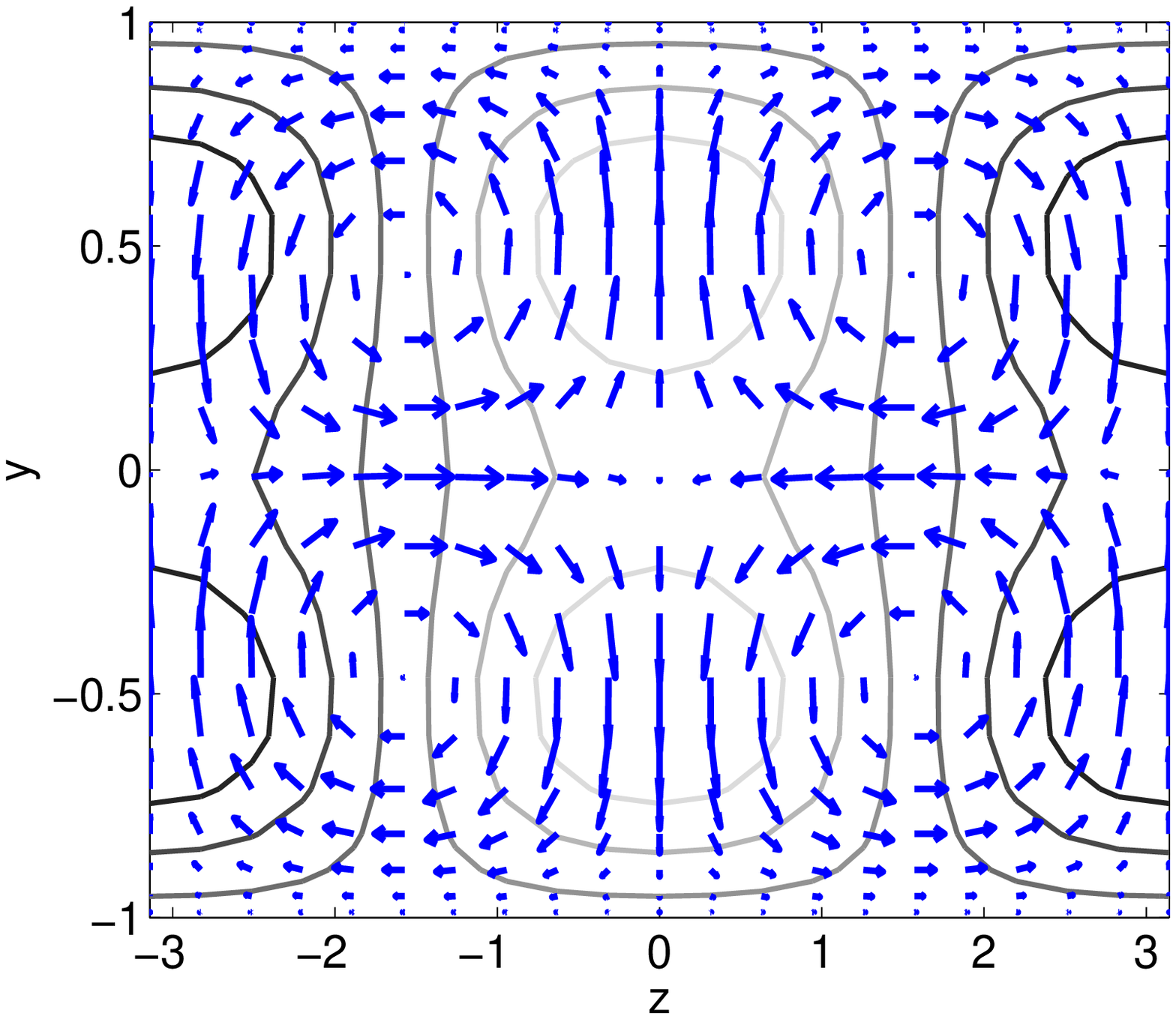}
  		\put(-3,80.5){$(c)$}
   	\end{overpic}  
   	\hspace{0.9cm}
   	\begin{overpic}[height=5cm,tics=10]{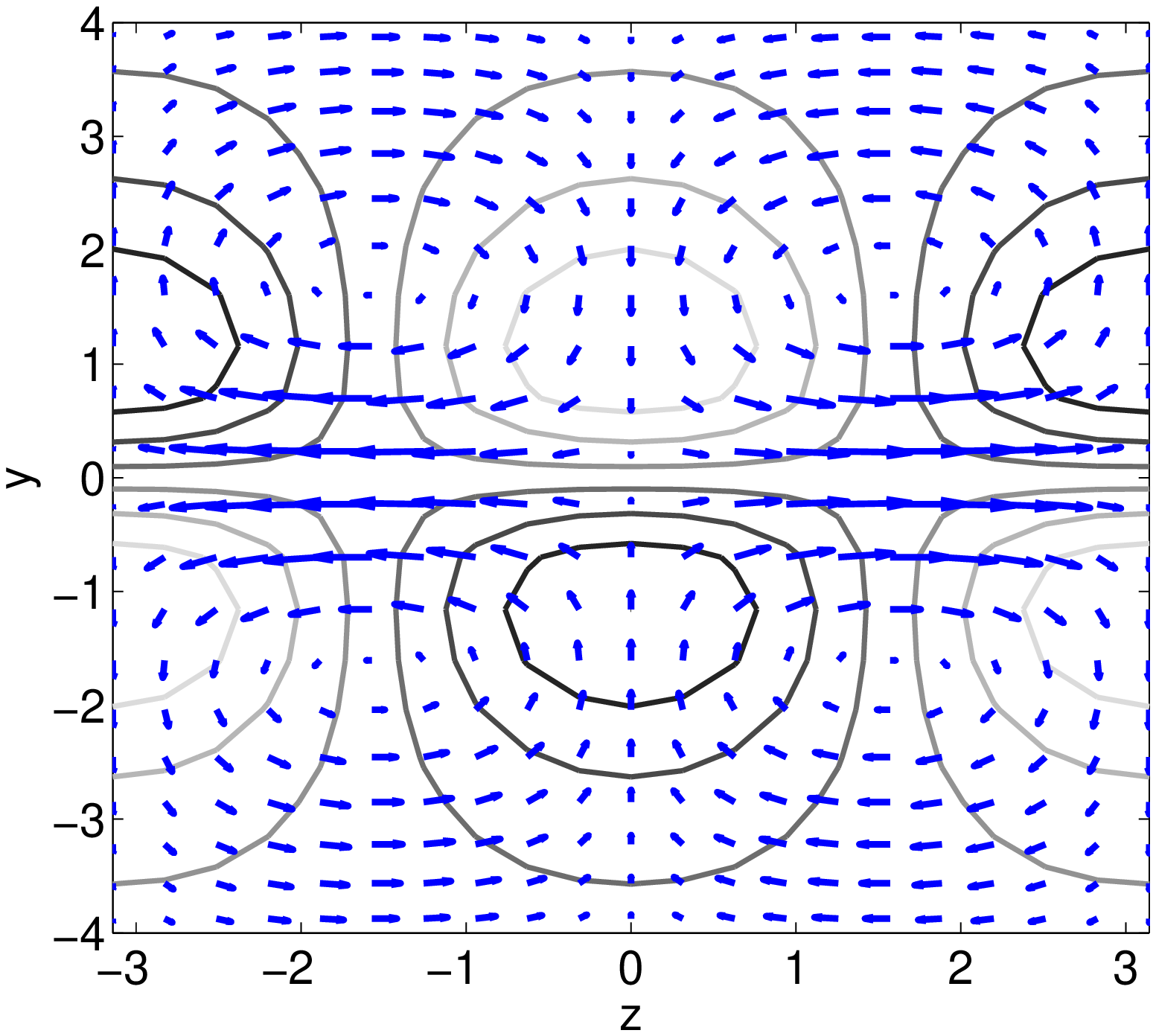}
  		\put(-3.5,82.5){$(d)$} 
   	\end{overpic} 
   } 
   \vspace{0.35cm}  
   \centerline{	
   \hspace{0.05cm}
  	\begin{overpic}[height=5cm,tics=10]{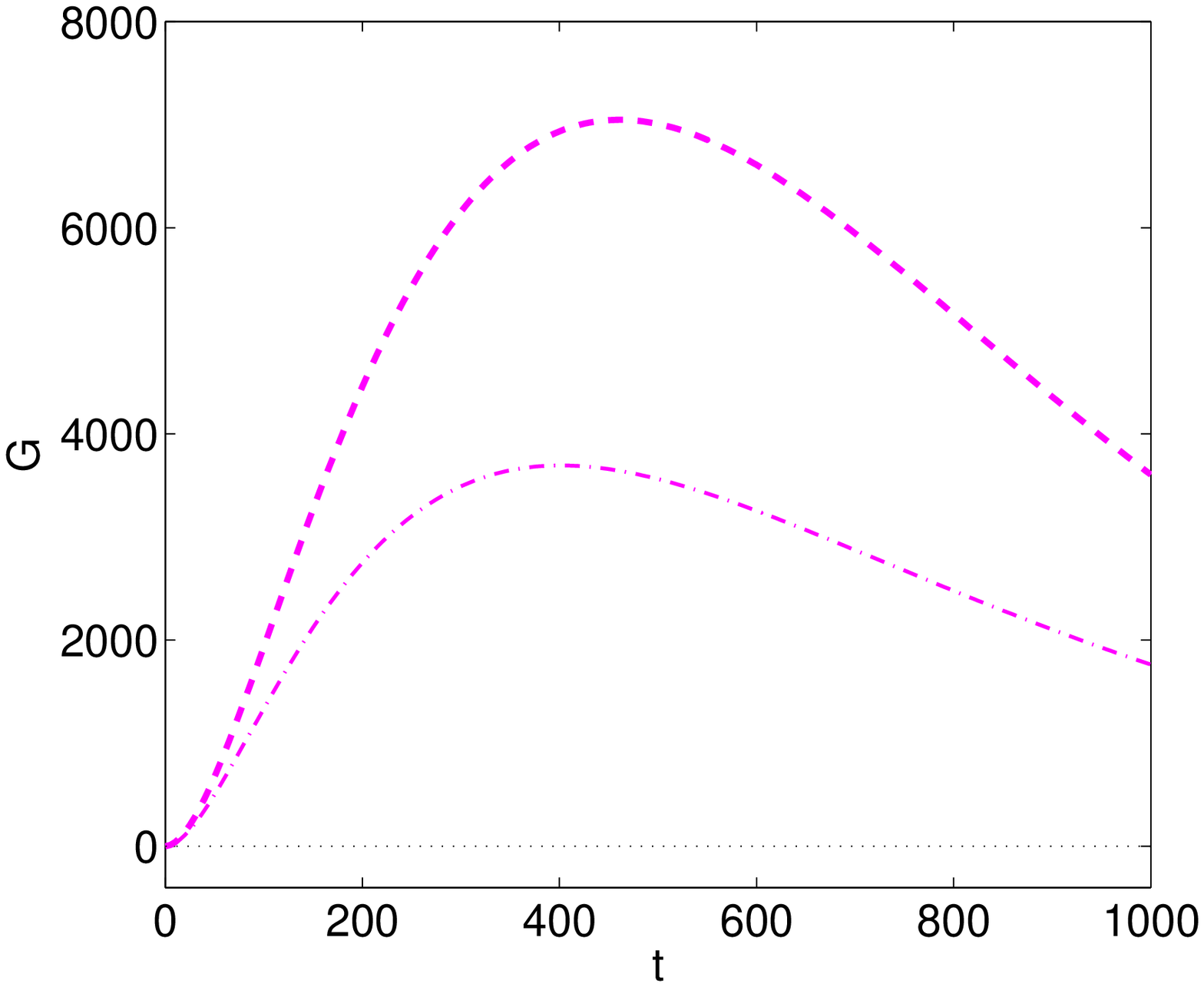}
  		\put(-4.5,75){$(e)$}
  		\put(71,65){\textcolor{magenta}{Optimal} } 
  		\put(32,30){\textcolor{magenta}{First suboptimal} }
   	\end{overpic}  
   	\hspace{0.45cm}
   	\begin{overpic}[height=5cm,tics=10]{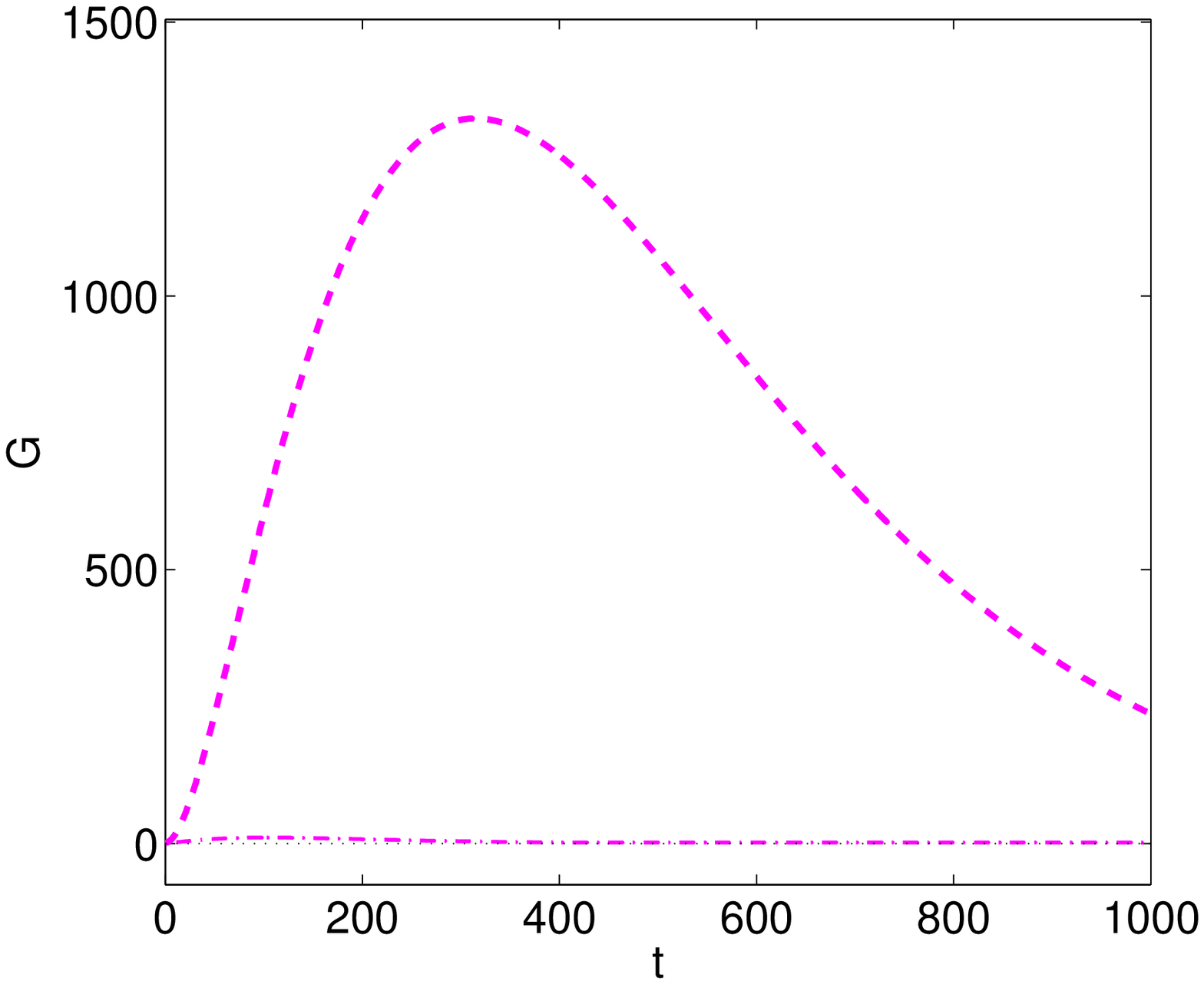}
  		\put(-5,75){$(f)$} 
  		\put(53,65){\textcolor{magenta}{Optimal} } 
  		\put(28,17){\textcolor{magenta}{First suboptimal} }
  	\end{overpic} 
   }  
   \caption{
   $(a),(b)$ Optimal and $(c),(d)$ first suboptimal perturbations for transient growth only.
   Arrows: initial $(v,w)$ components (streamwise vortices); 
   contours: $u$ component after optimal amplification (streamwise streaks, low speed in dark, high speed in light).
   $(e),(f)$ Growth versus time for  optimal and first suboptimal perturbations.
   $(a),(c),(e)$ Plane channel flow, $\Rey=6000$, $(\alpha,\beta)=(0,2)$;
   $(b),(d),(f)$ Mixing layer,       $\Rey=100$,  $(\alpha,\beta)=(0,0.35)$.   
   }
\label{fig:optTG}
\end{figure}

Optimally amplified perturbations in the plane channel flow and in the mixing layer are shown in figure~\ref{fig:optTG}$(a),(b)$ for oblique perturbation wavenumbers $(\alpha,\beta)=(0,2)$ and $(\alpha,\beta)=(0,0.35)$ respectively. They lead to the maximal optimal growth over all wavenumbers $\max_{\alpha,\beta} G_{opt}(T)=7047$ and $\max_{\alpha,\beta} G_{opt}(T)=1328$.
In both flows, streamwise vortices at $t=0$ (arrows, $v$ and $w$ components) are amplified into streamwise streaks at $t=T$ (contours, $u$ component).
The same computational method also yields a family of orthogonal suboptimal perturbations.

In the plane channel flow, the first suboptimal perturbation (shown in figure~\ref{fig:optTG}$(c)$) leads to a maximal amplification $G_{subopt}=3694$ (figure~\ref{fig:optTG}$(e)$). This is smaller than $G_{opt}(T)$ but of comparable order, while the amplification of following suboptimals is smaller by orders of magnitude, as already observed by \cite{Butler92}.
In the mixing layer, the first suboptimal perturbation  (shown in figure~\ref{fig:optTG}$(d)$) leads to a maximal amplification $G_{subopt}=16$,  two orders of magnitude smaller than $G_{opt}(T)$ (figure~\ref{fig:optTG}$(f)$).
\textcolor{black}{
In both flows, the typical time scale  for transient growth/decay is of the order of $10^2$.
}

If we now turn to \textit{combined} optimisation, we look for perturbations that, after amplification, lead to  maximal stabilisation.
We optimise for the \textit{combined}  stabilising effect $\ev_{2r}^c$ as
\begin{subeqnarray}
\min_{||\uu(0)||=1}  \ev_{2r}^c
& = &
 \min_{||\uu(0)||=1}  \ps{u(t)}{\frac{1}{2} \left( \widetilde\SS_{2r}+\widetilde\SS_{2r}^T \right) u(t)}
\\ 
& =& \min_{||\uu(0)||=1}  \ps{\evol_u(t) \uu(0)}{\frac{1}{2} \left( \widetilde\SS_{2r}+\widetilde\SS_{2r}^T \right)  \evol_u(t) \uu(0)}
\\
& =& \ev_{min} \left\{\frac{1}{2} \evol^\dag_u  \left( \widetilde\SS_{2r}+\widetilde\SS_{2r}^T \right) \evol_u \right\} 
\label{eq:comb}
\end{subeqnarray}
Here we  used $U_1=u(t)$ as the flow modification in (\ref{eq:mostdestab}) and neglected $v$ and $w$ components, which will be justified \textit{a posteriori}. Accordingly, $\evol_u$ is a shorthand notation for the action of $\evol$ followed by an extraction of the streamwise component.
The combined optimal perturbation is denoted $\uu_{opt}^c$.

\subsection{Results}

\begin{figure}
  \psfrag{t}[t][]{$t$}
  \psfrag{lam2r}[][][1][-90]{$\ev_{2r}^c \quad$}
  \vspace{0.5cm}
  \centerline{	
    \begin{overpic}[height=6.2cm,tics=10]{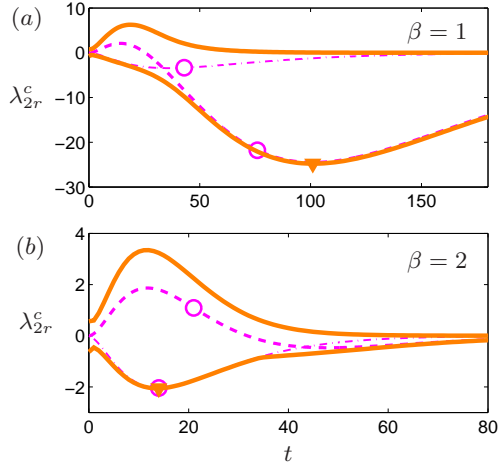}
	  	\put(-3,91){$(a)$} 
	  	\put(-2,43){$(b)$}
	  	\put(80,88){$\beta=1$}
	  	\put(80,40){$\beta=2$}
   	\end{overpic}   
  }    
   \caption{   
   Optimal \textit{combined} transient growth and eigenvalue variation in the mixing layer at $\Rey=100$.
   Thick lines show the optimal combined  effects for stabilisation and destabilisation;
   triangles indicate the maximal stabilising effect over all times.
  Dashed lines (resp. thin dash-dotted lines) show to 
 optimal (resp. first suboptimal) perturbations for transient growth only;
   circles: corresponding eigenvalue variation at optimal amplification time. 
Transient growth at 
   $(a)$~$(\alpha=0,\beta=1)$  and
   $(b)$~$(\alpha=0,\beta=2)$; 
   effect on the leading eigenvalue $(\alpha_0,\beta_0)=(0.45,0)$.    
   }
\label{fig:opt_lam_combined}
\end{figure}

Optimisation was carried out with transient growth at oblique wavenumbers $(\alpha,\beta)=(0,\beta)$, and with destabilisation/stabilisation of the most unstable mode $(\alpha_0,\beta_0)=(\alpha_{0,max},0)$.
\textcolor{black}{
In this section we focus on the mixing layer flow at $\Rey=100$, where the characteristic instability time scale $1/\ev_r$ is of the order of $10^0$, much smaller than the evolution time scale of the streaks. 
In the plane channel flow at $\Rey=6000$, the instability is weaker and the separation of time scales does not hold.
}
The evolution of combined optimal effects is shown as thick lines in figure~\ref{fig:opt_lam_combined}.
At $t=0$, the result of the optimisation is simply the  optimal flow modification without transient growth, as already presented in section 
\ref{sec:mixinglayer}. 
At later times, the combined optimisation identifies perturbations that achieve the best trade-off between amplification from 0 to $t$ and destabilisation/stabilisation at $t$.

Optimisation at $\beta=1$ 
yields a large stabilising effect over an extended time interval, thanks to the combination of amplification and flow modification. 
Indeed, combined optimal perturbations  are very similar to  transient growth-only optimal perturbations, whose effect (dashed line) is as large as the combined optimal effect.

At $\beta=2$, however, 
transient growth-only optimal perturbations  have a \textit{destabilising} effect and therefore cannot be used to control the flow. 
Nevertheless, a \textit{stabilising} combined effect can be obtained by choosing other perturbations, as indicated by the lower thick line.
The stabilising combined optimal perturbations turn out to be similar to the first suboptimal  perturbations for transient growth, which exploit the same lift-up mechanism but have the opposite symmetry in $y$: two vortices amplified into two streaks  on each side of the shear layer (fig.~\ref{fig:optTG}$(d)$), whereas optimal perturbations consist of one single vortex amplified into one streak  centred in $y=0$ (fig.~\ref{fig:optTG}$(b)$).

In all cases, we observed that the combined optimal perturbations quickly became and remained unidirectional: the $v_{opt}^c$ and $w_{opt}^c$ components were much smaller than $u_{opt}^c$ well before the time of largest destabilisation/stabilisation, which justifies (\ref{eq:comb}).

\begin{figure}
  \psfrag{b}[t][]{$\beta$}
  \psfrag{lam2r}[][][1][-90]{$\ev_{2r}^c$}
  \vspace{1cm}
  \centerline{	
   	\begin{overpic}[height=4.917cm,tics=10]{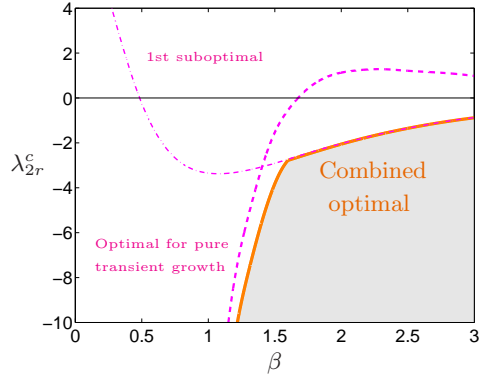}
	  	\put(17,25){\tiny \textcolor{magenta}{Optimal for pure}}	  	
	  	\put(17,20){\tiny \textcolor{magenta}{transient growth}}	  	
	  	\put(28,65){\tiny \textcolor{magenta}{1st suboptimal}}	  	
	  	\put(65,40){\textcolor{myorange}{Combined}} 
	  	\put(67,33){\textcolor{myorange}{optimal}} 
   	\end{overpic}  
  }    
   \caption{Largest stabilising effect $\ev_{2r}^c$ (thick line) obtained from the combined optimisation of transient growth and eigenvalue variation, versus control wavenumber $\beta$. 
Any other perturbation falls above the thick line, i.e. outside the grey region.
   The  dashed line (resp. thin dash-dotted line) shows the stabilising effect of transient growth-only optimal (resp. first suboptimal) perturbations.
Mixing layer, $\Rey=100$, $\alpha_0=\alpha_{0,max}=0.45$.}
\label{fig:lam2r_combined}
\end{figure}

Figure~\ref{fig:lam2r_combined} summarises the variation of the maximal combined stabilising effect with $\beta$. 
The thick line shows the optimum over all times, and therefore provides a bound for the eigenvalue variation.
At small wavenumbers $\beta \lesssim 1.5$, the effect of optimally amplified streaks closely follows the optimal combined stabilising effect.
Combined optimal perturbations are very similar to optimally amplified perturbations, whose effect (dashed line) is of the same order, and is reached after a comparable amplification time.
In this range of wavenumbers, the optimal strategy to stabilise the flow at minimal cost is therefore precisely to introduce optimally amplified streaks.
Other types of perturbations do not perform as well, either because they undergo a smaller amplification or because the flow modification that they induce has a smaller stabilising effect.
At larger wavenumbers $\beta \gtrsim 1.5$, optimally amplified perturbations are destabilising. Instead, the largest combined stabilisation is obtained with the first suboptimal perturbations for transient growth (dash-dotted line), which exhibit smaller  cross-stream structures but undergo the same lift-up mechanism.

\subsection{Discussion}

In previous sections we have presented optimally stabilising flow modifications $U_1$, optimal perturbations $u_{opt}$ for transient growth only, and combined optimal perturbations $u_{opt}^c$ for transient growth and stabilisation. 
We now compare the effect of these 2D (spanwise-periodic) modulations with the effect of 1D (spanwise-invariant) base flow modifications.
Since the effect of spanwise-periodic modifications $\epsilon U_1(y) \cos(\beta z)$ is quadratic, 
$|\ev-\ev_0| \sim \epsilon^2$, one can expect that it should be overcome at small amplitudes $\epsilon$ by the non-zero linear effect 
$|\ev-\ev_0| \sim \epsilon$ of spanwise-invariant modifications $\epsilon U_1^{1D}(y)$.
We use  first-order sensitivity \citep{Bottaro03} to compute the eigenvalue sensitivity to 1D flow modification
$S_1 
= \ex\bcdot(-\uua_0 \bcdot \bnabla \uu_0^H + \overline\uu_0 \bcdot \bnabla \uua_0) 
= i\alpha_0 (2\overline u_0 \ua_0 + \overline v_0 \va_0) + \overline v_0 \partial_y \ua_0$.
We then deduce the optimal 1D flow modification, equal to the real part of the sensitivity itself, $U_1^{1D}=S_{1r}$, and the maximal growth rate variation, $\max\ev_{1r}=\ps{S_{1r}}{U_1^{1D}}=\ps{S_{1r}}{S_{1r}}$.

\begin{figure}
  \psfrag{eps}[t][]{$\epsilon$}
  \psfrag{real(lambda0)}[][][1][-90]{$\ev_{r}\quad$}
  \psfrag{dreal(lambda0)}[][][1][-90]{$\delta\ev_{r}\quad$}
  \psfrag{y}[][][1][-90]{$y$}
  \psfrag{q1D}[t][]{$U_1^{1D}$}
  \psfrag{U1}[t][]{$U_1^{2D}$}
  \psfrag{qcomb(T)}[t][]{$u_{opt}^c(T)$} 
  \vspace{0.5cm} 
  \centerline{
    \begin{overpic}[height=5cm,tics=10]{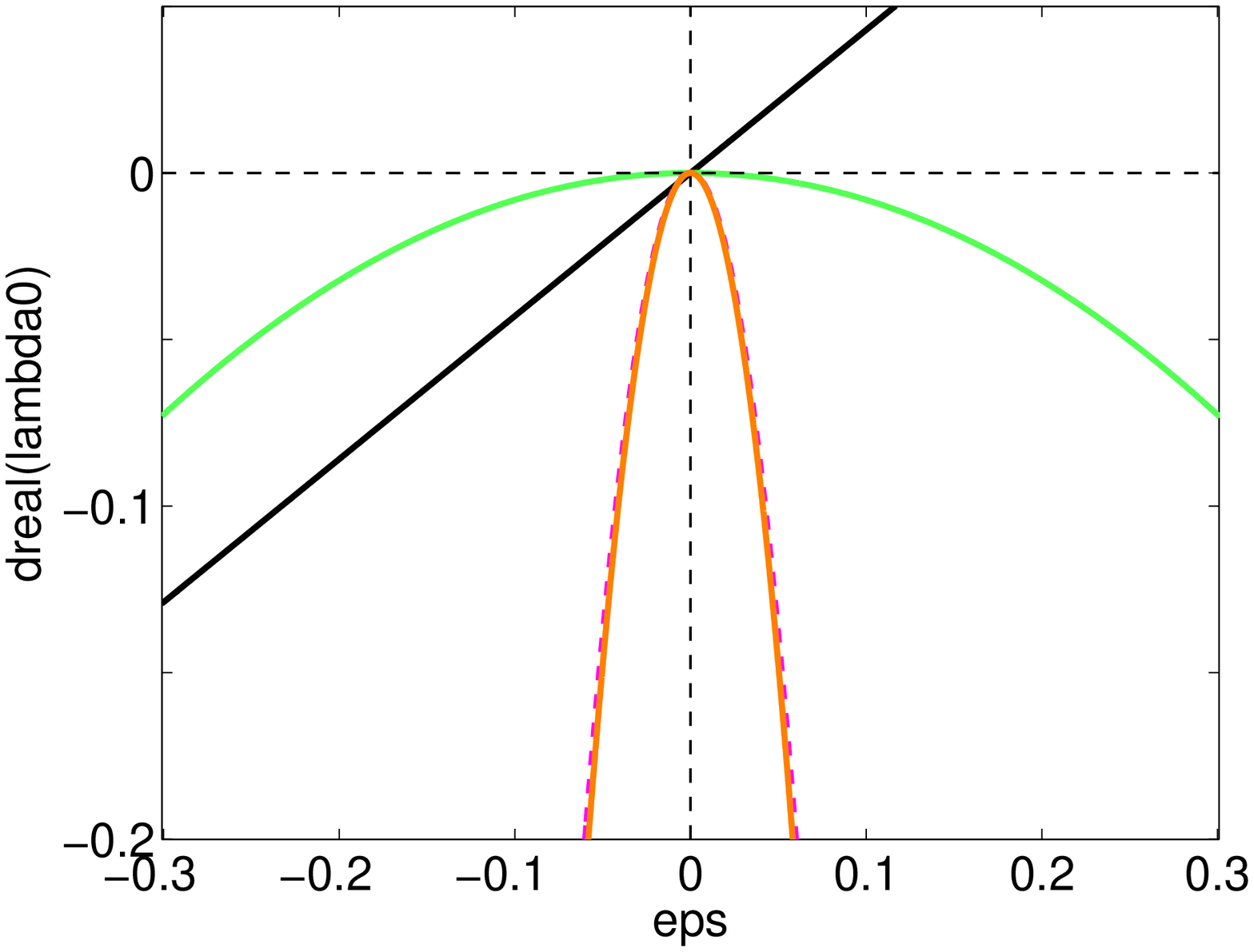}
	  	\put(-3,70){$(a)$}
	  	\put(72,47)  { \textcolor{greenstab}{$U_1^{2D}$} }	
	  	\put(24,32)  { $U_1^{1D}$ }
	 	\put(63,25)  { \textcolor{myorange}{$u_{opt}^c(T)$,} }
	  	\put(63,18)  { \textcolor{magenta}{$u_{opt}(T)$} }		  	
   	\end{overpic}  
	\hspace{0.2cm}
	\begin{overpic}[height=5.cm,tics=10]{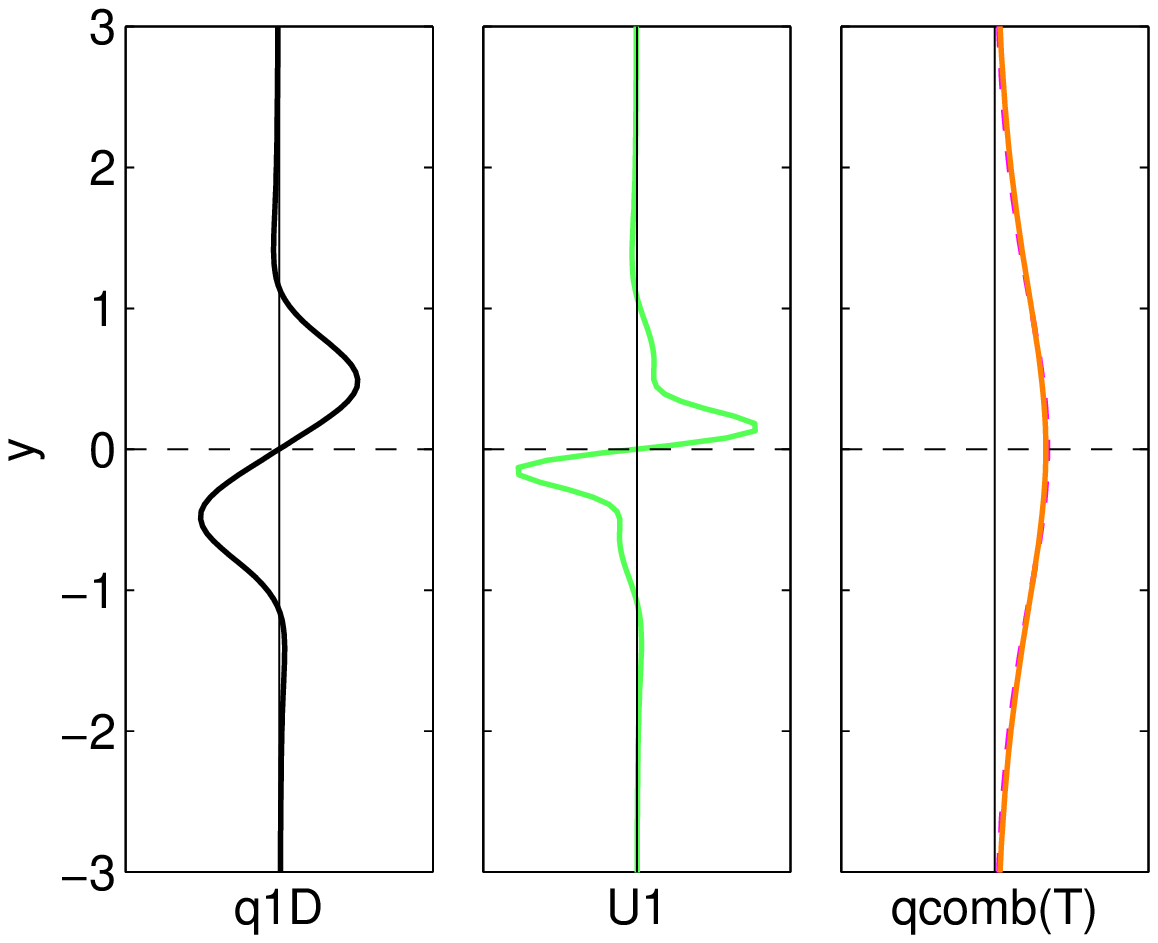}
	  	\put(1,74){$(b)$}	
   	\end{overpic}  
  }   
  \vspace{0.2cm}
  \caption{$(a)$ Growth rate variation for \textit{combined} optimal perturbations as a function of amplitude $\epsilon$, compared to 1D and 2D optimal flow modifications for pure stabilisation.
   $(b)$ Corresponding profiles of streamwise velocity: 
   1D optimal $U_1^{1D}$,  
   2D optimal $U_1^{2D}$, and 
   \textit{combined} optimal after amplification $u_{opt}^c(T)$. 
   The optimal perturbation for pure transient growth $u_{opt}(T)$ is also shown as a dashed line superimposed onto the \textit{combined} optimal, but the two curves are indistinguishable.
   Mixing layer, $\beta=0.8$, eigenmode $(\alpha_0,\beta_0)=(0.45,0)$, $\Rey=100$.
   }
\label{fig:POIS_compare1D-with_TG}
\end{figure}

We compare in figure \ref{fig:POIS_compare1D-with_TG}$(a)$  the effects of optimal 1D and 2D flow modifications of unit norm \textit{per unit spanwise length}, taking into account the total cost of modifying the flow over a given $z$ region. 
For intermediate wavenumber values (here $\beta=0.8$) and for the amplitudes of interest,  the optimal 1D flow modification $U_1^{1D}$ has a stronger stabilising effect than its 2D counterpart $U_1$ (also denoted $U_1^{2D}$  for clarity). 
Remarkably, combined optimal perturbations $u_{opt}^c$ have an even larger stabilising effect, as a result of  transient growth. 
As mentioned earlier, $u_{opt}$ has an effect very similar to that of $u_{opt}^c$ for the wavenumber considered here.

Figure \ref{fig:POIS_compare1D-with_TG}$(b)$ shows the corresponding velocity profiles.
The optimal 1D modification appears to be qualitatively similar to the optimally stabilising 2D modification: 
$U_1^{1D}$  has antisymmetric structures localised on each side of the shear layer but farther away than $U_1^{2D}$ (in a fashion actually similar to the optimally destabilising 2D modification at small wavenumber; see fig.~\ref{fig:TANH_U0U1branchI}).
It should be noted that for  1D spanwise-invariant  flow modification it is enough to change the sign of $U_1^{1D}$ to turn a stabilising effect into a destabilising one and vice-versa, in contrast to 2D spanwise-periodic flow modification since stabilising and destabilising $U_1^{2D}$ have different structures (figures~\ref{fig:POIS_U0} and \ref{fig:TANH_U0U1branchI}).

Combined optimal perturbations $u_{opt}^c$ have radically different velocity profiles, more widely distributed than $U_1$ and with the opposite symmetry.
This is the result of initial perturbations being
smoothed out by diffusion during transient amplification; one can therefore expect that these optimal structures would be easier to produce robustly in experiments. 
Again, it also indicates that combined optimal perturbations have an   intrinsic stabilising effect that is suboptimal when considered alone, but they achieve a large combined stabilising effect when taking advantage of amplification.

\section{Conclusion}

We have  determined analytically the second-order sensitivity of the leading eigenvalue in  parallel flows
with respect to small-amplitude spanwise-periodic velocity modifications  of wavenumber $\beta$.
The explicit derivation of a second-order sensitivity operator allowed us to obtain the eigenvalue variation induced by any such  flow modification without ever solving the modified eigenvalue problem or calculating the first-order eigenmode correction.
Predictions from sensitivity analysis have been validated against numerical calculations of the full stability problem, and  quadratic variations with modification amplitude were observed.

For any pair of eigenmode streamwise wavenumber $\alpha_0$ and base flow modification spanwise wavenumber $\beta$, we have maximised the eigenvalue variation and determined the most destabilising and stabilising flow modifications. 
In the plane channel flow, the optimal modifications are localised close to the walls near critical layers, and show little variation with  $\beta$; 
in the mixing layer, they are centred around the location of maximum shear, and become more concentrated with increasing $\beta$.

We observed that the optimal variations in growth rate  increased like $\ev_{2r}\sim\beta^{-2}$ for small control wavenumbers, and explained that this scaling was consistent with an interaction between unperturbed 2D and 3D eigenmodes.
Spanwise-periodic flow modifications also appeared to have a larger effect  on modes of larger streamwise wavenumber  $\alpha_0$.
Modifications optimised for the most unstable mode 
have a robust effect on modes with other $\alpha_0$. A splitting of 3D modes with $\beta_0=\pm\beta$ and $\beta_0=\pm\beta/2$ suggests the choice of large enough control wavenumbers so as not to destabilise weakly damped 3D modes.

In a second step aimed at increasing the stabilising effect and thus reducing the required control amplitude, we optimised simultaneously the 
\textcolor{black}{linear} transient growth of initial perturbations and the subsequent eigenvalue variation induced by the resulting flow modification,
\textcolor{black}{under the assumption of separation of time scales between transient growth and instability.}
In the mixing layer, this combined optimisation revealed that the perturbations undergoing the largest transient growth were also the combined optimals for small enough control wavenumbers, $\beta \lesssim 1.5$, which justifies \textit{a posteriori} the ``vaccination'' strategy proposed by \cite{Cossu02} to take advantage of the amplification of streamwise vortices into streamwise streaks through lift-up. 
As  $\beta$  increases, optimally amplified perturbations gradually become  destabilising, and the combined optimal turns out to correspond to the first suboptimal for pure transient growth.

The method readily applies to the quadratic sensitivity of the eigenmode frequency, if one wishes, for instance, to detune vortex the shedding frequency, and can easily be extended to other flow quantities.

Combined optimal perturbations could be created experimentally with a method similar to that of \cite{Fra05}, who used roughness elements in a boundary-layer flow to produce streamwise vortices amplified by transient growth into streamwise streaks. 
In the mixing layer these roughness elements  could be placed on one or both sides  of the splitting plate used to create the shear layer. 
One could also imagine using wall actuation to create the initial streamwise vortices.

We are presently working on generalising  the second-order optimisation technique presented in this paper to non-parallel flows.
In spatially developing flows, the method is conceptually similar but involves a number of complications. 
First, the large number of degrees of freedom prevents the explicit calculation of the inverse operator $(\ev_0\EE+\AAA_0)^{-1}$, but optimisation should still be possible since it only requires repeated evaluations of matrix-vector products.
Second, the derivation of adjoint operators is technically more involved.
In view of recent publications which describe the quadratic dependence of the growth rate on the amplitude of spanwise-periodic control in spatially developing flows such as wakes behind a circular cylinder \citep{Kim05,DelGuercio2014} and behind a flat plate of finite thickness \citep{Tammi14}, this generalisation appears to be  a promising research direction.

\appendix
\section{Stability operators}
\label{sec:app_stab}

We write explicitly the eigenvalue problem  
(\ref{eq:LNS_evp}) 
for perturbations 
$\qq(y,z)\exp(i\alpha_0 x+\ev t)$ linearised around 
the  base flow $U(y,z)\ex$:
\begin{subeqnarray}
&& \lambda u + i \alpha_0 u U  +  v \partial_y U + w \partial_z U +i\alpha_0 p -\Rey^{-1} \nabla_{\alpha_0} u =0
\\
&& \lambda v + i\alpha_0 v U  + \partial_y p  -\Rey^{-1} \nabla_{\alpha_0} v = 0 
\\
&& \lambda w + i\alpha_0 w U   + \partial_z p -\Rey^{-1} \nabla_{\alpha_0} w  = 0 
\\ 
&& i\alpha_0 u+ \partial_y v + \partial_z w =0,
\end{subeqnarray}
where $\nabla_{\alpha_0} = -\alpha_0^2+\partial_{yy} + \partial_{zz}$.

From the expansion 
(\ref{eq:exp}), we find at leading order ($\epsilon^0$) 
the eigenvalue problem 
(\ref{eq:evp1111})
 $(\lambda_0 {\bf E}+{\bf A}_0) {\bf q}_0=\bf 0$, where
\begin{subeqnarray}
& {\bf E}=\left[\begin{array}{cccc}
1 & 0 & 0 & 0 \\
0 & 1 & 0 & 0 \\
0 & 0 & 1 & 0 \\
0 & 0 & 0 & 0 
\end{array}\right],
\\
& {\bf A}_0=\left[\begin{array}{cccc}
i\alpha_0 U_0-\Rey^{-1} \nabla_{\alpha_0} & \partial_y U_0 & 0 & i\alpha_0\\
0 & i\alpha_0 U_0-\Rey^{-1} \nabla_{\alpha_0} & 0 & \partial_y \\
0 & 0 & i\alpha_0 U_0-\Rey^{-1} \nabla_{\alpha_0} &  \partial_z\\
i \alpha_0  & \partial_y & \partial_z & 0 
\end{array}\right].
\end{subeqnarray}

At first order ($\epsilon^1$)
we find equation (\ref{eq:evp1}) 
for the first-order eigenmode correction $\qq_1$, 
i.e. $(\lambda_0 {\bf E} +{\bf A}_0){\bf q}_1 + (\lambda_1 {\bf E} +{\bf A}_1){\bf q}_0 = \bf 0$, 
where
\begin{equation}
{\bf A}_1=\left[\begin{array}{cccc}
i\alpha_0 U_1\cos(\beta z) & \partial_y U_1 \cos(\beta z) & -\beta U_1\sin(\beta z) & 0\\
0 & i\alpha_0 U_1\cos(\beta z) & 0 & 0 \\
0 & 0 & i\alpha_0 U_1\cos(\beta z) &  0\\
0 & 0 & 0 & 0 
\end{array}\right].
\end{equation}

\section{Sensitivity operators}
\label{sec:app_sensit}

The second-order eigenvalue variation 
(\ref{eq:ev2})
 is manipulated in order to separate the base flow modification 
  and the second-order sensitivity operator: 
$\ev_2=\pps{U_1}{\SS_2 U_1}$.
To this aim, we introduce the adjoint operator $\AAA_1^\dag$ such that 
\be \ev_2 
 = \pps{\qq_0^\dag}{\AAA_1 (\ev_0\EE+\AAA_0)^{-1}\AAA_1\qq_0}
 = \pps{\AAA_1^\dag \qq_0^\dag}{ (\ev_0\EE+\AAA_0)^{-1}\AAA_1\qq_0},\ee
 and we isolate the flow modification by rewriting
$\AAA_1^\dag\qqa_0=\MM U_1$
and 
$\AAA_1\qq_0=\LL U_1$, where
\be
\MM
= \left[
\begin{array}{c}
- i \alpha_0   u_0^\dag \cos(\beta z) \\ 
   \left( u_0^\dag \partial_y - i \alpha_0   v_0^\dag \right)  \cos(\beta z)  \\ 
- \beta u_0^\dag \sin(\beta z)  \\ 
0 
\end{array}  \right],
\quad
\LL
= \left[
\begin{array}{c}
\left( i \alpha_0 u_0 +  v_0 \partial_y \right) \cos(\beta z) \\ 
  i \alpha_0 v_0 \cos(\beta z)  \\ 
0 \\ 
0
\end{array}  \right].
\ee
Finally, the second-order eigenvalue variation reads 
\be \ev_2=\pps{\MM U_1}{ (\ev_0\EE+\AAA_0)^{-1}\LL U_1}
=\pps{ U_1}{ \MM^\dag(\ev_0\EE+\AAA_0)^{-1}\LL U_1},
\ee
where
\be
\MM^\dag
 = \left[
 i \alpha_0  \overline u_0^\dag \cos(\beta z),\, 
\left(
   i \alpha_0\overline v_0^\dag 
  -\partial_y\overline u_0^\dag
  -\overline u_0^\dag\partial_y 
  \right)\cos(\beta z)   ,\,  
-\beta \overline u_0^\dag  \sin(\beta z),\, 0 
\right],
\ee
and we identify the second-order sensitivity operator $\SS_2=\MM^\dag(\ev_0\EE+\AAA_0)^{-1}\LL$.

\section{Transformation to spanwise-independent operators}
\label{sec:app_z}

We give details about the simplification of the second-order sensitivity operator, from the general expression $\SS_2$ given in (\ref{eq:ev21}) 
to the $z$-independent form  $\widetilde\SS_2$ given in 
(\ref{eq:ev23}),
which is made possible by the explicit expressions of $\MM^\dag$ and $\LL$ being available.

We denote $\aa=(a_u,a_v,a_w,a_p)^T$ a solution of 
$ (\ev_0\EE+\AAA_0) \aa = \LL U_1 $:
\begin{subeqnarray}
&& (\ev_0+i\alpha_0 U_0-\Rey^{-1} \nabla_{\alpha_0})  a_u + \partial_y U_0 a_v + i\alpha_0 a_p = \cos(\beta z) \left( i \alpha_0 u_0 +  v_0 \partial_y \right) U_1
\\
&& (\ev_0+ i\alpha_0 U_0-\Rey^{-1}\nabla_{\alpha_0})  a_v +\partial_y a_p = \cos(\beta z) i \alpha_0 v_0 U_1
\\
&& (\ev_0+ i\alpha_0 U_0-\Rey^{-1} \nabla_{\alpha_0}) a_w +  \partial_z a_p = 0
\\
&& i \alpha_0 a_u + \partial_y a_v + \partial_z a_w = 0.
\end{subeqnarray}
A close inspection reveals that  this solution is necessarily of the form 
\be 
\aa = (\widetilde a_u \cos(\beta z), \widetilde a_v \cos(\beta z), \widetilde a_w \sin(\beta z), \widetilde a_p \cos(\beta z))^T.
\ee
Therefore, the  second-order eigenvalue variation reads
\begin{subeqnarray}
\ev_2 
&=& \pps{U_1}{\SS_2 U_1}
= \pps{U_1}{\MM^\dag(\ev_0\EE+\AAA_0)^{-1}\LL U_1}
= \pps{U_1}{\MM^\dag \aa}
\\
&=& 
\pps{U_1}{\left(i \alpha_0  \overline u_0^\dag \widetilde a_u
+
\left(i \alpha_0\overline v_0^\dag 
  -\partial_y\overline u_0^\dag
  -\overline u_0^\dag\partial_y 
  \right) \widetilde a_v
  \right) \cos^2(\beta z)   
-\beta \overline u_0^\dag \widetilde a_w \sin^2(\beta z)} \qquad
\\
&=& 
\frac{1}{2}
\ps{U_1}{\left(i \alpha_0  \overline u_0^\dag \widetilde a_u
+
\left(i \alpha_0\overline v_0^\dag 
  -\partial_y\overline u_0^\dag
  -\overline u_0^\dag\partial_y 
  \right) \widetilde a_v
  \right)   
- \beta \overline u_0^\dag \widetilde a_w }
\end{subeqnarray}
where the last equality comes from the integrals of $\cos^2(\beta z)$ and 
$\sin^2(\beta z)$ over one spanwise wavelength being equal to $\pi/\beta$, 
yielding $1/2$ when normalised by the  wavelength.
Finally the second-order sensitivity operator becomes independent of $z$:
\be
 \ev_2  = \ps{ U_1}{\frac{1}{2}\widetilde\MM^\dag(\ev_0\EE+\widetilde\AAA_0)^{-1} \widetilde\LL U_1},
\ee
where
\be
\widetilde
\MM^\dag
 = \left[
 i \alpha_0  \overline u_0^\dag,\,
   i \alpha_0\overline v_0^\dag 
  -\partial_y\overline u_0^\dag
  -\overline u_0^\dag\partial_y,\,
-\beta \overline u_0^\dag,\, 
0 
\right],
\quad
\widetilde \LL
= \left[
\begin{array}{c}
 i \alpha_0 u_0 +  v_0 \partial_y   \\ 
  i \alpha_0 v_0 \\ 
0 \\ 
0
\end{array}  \right],
\ee
\begin{equation}
\widetilde{\bf A}_0
=
\left[\begin{array}{cccc}
i\alpha_0 U_0-\Rey^{-1} \nabla_{\alpha_0\beta} & \partial_y U_0 & 0 & i\alpha_0\\
0 & i\alpha_0 U_0-\Rey^{-1}  \nabla_{\alpha_0\beta} & 0 & \partial_y \\
0 & 0 & i\alpha_0 U_0-\Rey^{-1}  \nabla_{\alpha_0\beta} &  -\beta\\
i \alpha_0  & \partial_y & \beta & 0 
\end{array}\right],
\end{equation}
and  $\nabla_{\alpha_0\beta} = -\alpha_0^2+\partial_{yy} -\beta^2$.

\section{Scaling in the small-$\beta$ limit}
\label{sec:app_scaling}

Consider the unperturbed base flow and 
the corresponding unstable 2D eigenmode $\qq_0$ solution of 
the eigenvalue problem
$(\lambda_0 {\bf E}+{\bf A}_0) {\bf q}_0=\bf 0$.
There also exist 3D eigenmodes of spanwise wavenumber $\beta_0\neq 0$, which are all less unstable.
For small values of the spanwise wavenumber $|\beta_0| \ll 1$, the eigenvalue varies like $\ev(\beta_0)\simeq\ev_0-C \beta_0^2$,
$C>0$, as outlined in the following.
We estimate the eigenvalue variation for a small change of spanwise wavenumber from 
$\beta_0=0$ to 
$\beta_0 \ll 1$
by using the classical expression of first-order eigenvalue variation
$ \delta \ev =  - \ps{ {\bf q}_0^\dag } {\bdelta{\bf A}_0{\bf q}_0} 
/ 
\ps{ {\bf q}_0^\dag } {\EE {\bf q}_0}$,
where the variation of the linearised Navier-Stokes operator in this case is 
given by 
\begin{equation}
\bdelta\AAA_0
=
\left[\begin{array}{cccc}
\Rey^{-1}  \beta_0^2 & 0 & 0 & 0\\
0 & \Rey^{-1}  \beta_0^2 & 0 & 0 \\
0 & 0 & \Rey^{-1} \ \beta_0^2 &  i \beta_0\\
0 & 0 & i \beta_0 & 0 
\end{array}\right],
\end{equation}
and therefore $\delta \ev = \Rey^{-1}  \beta_0^2$.

When the base flow is perturbed with a spanwise-periodic modification of long wavelength (small $\beta$), there is interaction between the unperturbed 2D eigenmode of interest ($\beta_0=0$) and the unperturbed 3D eigenmodes of spanwise wavenumber $\pm\beta$, as described by \cite{Tammi14}; 
the resonance effect these authors proposed to explain large second-order sensitivity at small $\beta$ therefore leads in the present case to the second-order eigenvalue variation $|\ev_2| \sim |\delta\ev|^{-1} \sim \beta^{-2}$, as observed in
section~\ref{sec:POIS_optimflowmodif}.

\bibliographystyle{jfm}
\bibliography{2ndorder_biblio}

\end{document}